\documentclass[preprint]{aastex}

\usepackage{graphicx}
\usepackage[breaklinks,colorlinks,citecolor=blue,linkcolor=magenta]{hyperref}
\usepackage{natbib}
\usepackage{xcolor}
\usepackage{lscape}

\newcommand{\Msun}{\mbox{$M_{\odot}$}}
\newcommand{\msun}{\mbox{$M_{\odot}$}}

\newcommand{\Teff}{\mbox{$T_{\rm eff}$}}

\newcommand{\feh}{\mbox{[Fe/H]}}
\newcommand{\narratio}{\mbox{$N_{\rm TP-AGB}/N_{\rm RGB}$}}

\newcommand{\narratiot}{\mbox{$\frac{N_{\rm TP-AGB}}{N_{\rm RGB}}$}}

\defcitealias{Girardi2010}{G10}
\defcitealias{Melbourne2012}{M12}
\defcitealias{Rosenfield2014}{R14}

\begin{document}
\shorttitle{Evolution of TP-AGB Stars V}
\title{Evolution of Thermally Pulsing Asymptotic Giant Branch Stars V: Constraining the Mass Loss and Lifetimes of Intermediate Mass, Low Metallicity AGB Stars.\footnote{Based on observations made with the NASA/ESA Hubble Space Telescope, obtained from the Data Archive at the Space Telescope Science Institute, which is operated by the Association of Universities for Research in Astronomy, Inc., under NASA contract NAS 5-26555.}}

\shortauthors{P. Rosenfield et al.}

\author{Philip Rosenfield\altaffilmark{1}}
\affil{Harvard-Smithsonian Center for Astrophysics, 60 Garden St., Cambridge, MA 02138, USA}
\affil{Department of Physics and Astronomy G. Galilei, University of Padova, Vicolo dell'Osservatorio 3, I-35122 Padova, Italy}
\author{Paola Marigo}
\affil{Department of Physics and Astronomy G. Galilei, University of Padova, Vicolo dell'Osservatorio 3, I-35122 Padova, Italy}
\author{L\'eo Girardi}
\affil{Osservatorio Astronomico di Padova -- INAF, Vicolo dell'Osservatorio 5, I-35122 Padova, Italy}
\author{Julianne J.\ Dalcanton}
\affil{Department of Astronomy, University of Washington, Box 351580, Seattle, WA 98195, USA}
\author{Alessandro Bressan}
\affil{Astrophysics Sector, SISSA, Via Bonomea 265, I-34136 Trieste, Italy}
\author{Benjamin F.\ Williams}
\affil{Department of Astronomy, University of Washington, Box 351580, Seattle, WA 98195, USA}
\author{Andrew Dolphin}
\affil{Raytheon Company, 1151 East Hermans Road, Tucson, AZ 85756, USA}
\altaffiltext{1}{NSF Astronomy and Astrophysics Postdoctoral Fellow.}

\begin{abstract}
Thermally-Pulsing Asymptotic Giant Branch (TP-AGB) stars are relatively short lived (less than a few Myr), yet their cool effective temperatures, high luminosities, efficient mass-loss and dust production can dramatically effect the chemical enrichment histories and the spectral energy distributions of their host galaxies. The ability to accurately model TP-AGB stars is critical to the interpretation of the integrated light of distant galaxies, especially in redder wavelengths. We continue previous efforts to constrain the evolution and lifetimes of TP-AGB stars by modeling their underlying stellar populations. Using {\it Hubble Space Telescope (HST)} optical and near-infrared photometry taken of 12 fields of 10 nearby galaxies imaged via the ACS Nearby Galaxy Survey Treasury and the near-infrared HST/SNAP follow-up campaign, we compare the model and observed TP-AGB luminosity functions as well as the number ratio of TP-AGB to red giant branch stars. We confirm the best-fitting mass-loss prescription, introduced by \citet{Rosenfield2014}, in which two different wind regimes are active during the TP-AGB, significantly improves models of many galaxies that show evidence of recent star formation. This study extends previous efforts to constrain TP-AGB lifetimes to metallicities ranging $-1.59 \lesssim \feh \lesssim -0.56$ and initial TP-AGB  masses up to $\sim 4 M_\odot$, which include TP-AGB stars that undergo hot-bottom burning.
\end{abstract}


\section{Introduction}

Thermally Pulsing Asymptotic Giant Branch (TP-AGB) stars are fascinating in their structural complexity, their short lifetimes, and the dramatic impacts they have on their host galaxies. TP-AGB stars are known as important sources of dust \citep[e.g.,][]{Gail1985, Gail1999, Gehrz1989, Boyer2012, Melbourne2013}, they chemically enrich their ISM \citep[e.g.,][]{Marigo2001, Ventura2001, Karakas2007}, and they are critical to the interpretation of the spectral energy distributions of high-redshift galaxies \citep[e.g.,][]{Bertoldi2003, Valiante2009, Conroy2009, Melbourne2012}.

A full comprehension of the TP-AGB phase and its effects on its environment is critical to understanding galaxies' near and far infrared light output. However, TP-AGB stellar evolution is complex, with double shell fusion, large convective events (third dredge up), fusion at the base of the convective envelope (hot-bottom burning; HBB, for initial TP-AGB masses $\gtrsim 3\Msun$), and pulsations that help to create dust and eventually drive the mass loss rate.
As a consequence, even its most basic evolutionary property -- the TP-AGB lifetime -- differs greatly from set to set of evolutionary calculations \citep[which can be appreciated in Figure 3 of][]{Girardi2007}.

The strongest observational constraints on TP-AGB lifetimes first came from resolved stellar populations in the Magellanic Clouds \citep[MCs;][]{Frogel1990, vanLoon2005, Girardi2007} and then nearby galaxies \citep[][]{Girardi2007, Rosenfield2014}. The MCs clusters are natural laboratories for studying the TP-AGB phase as there are many clusters with well defined main sequence turn-offs \citep[e.g.,][]{Gallagher1996,Holtzman1999,Geha1998,Olsen1999}, known distances \citep[e.g.,][]{Pietrzynski2013}, deep optical and IR observations \citep[][]{Meixner2006, Blum2006}, and known metallicities \citep[e.g.,][]{Bica1998,Ferraro2006,Dirsch2000,Hill2000,Cole2005,DaCosta1998,Geisler2003}. However, individual MC clusters lack a statistically significant population of TP-AGB stars, which has lead some researchers to combine information from MC clusters based on age and derive TP-AGB lifetime constraints from ensemble populations \citep[e.g.,][]{Frogel1990, Girardi2007, Noel2013}. This method is now seen as problematic given many MC clusters happen to be of an age that temporarily boosts the number counts of TP-AGB stars, falsely signifying a longer lifetime. However, TP-AGB ``boosting'' does not eliminate MC clusters or fields from constraining the TP-AGB, it requires an extra calibration step to determine the amount of boosting when modeling stellar populations of $\sim 1.6$~Gyr \citep[see][]{Girardi2013}.

Outside the MCs, star clusters and fields containing significant numbers of TP-AGB stars are rarely be observed down to the old main sequence turn-off; however, modern color-magnitude diagram (CMD) fitting techniques allow their star formation histories (SFHs) to be well constrained up to distances of about 4~Mpc for high-resolution {\it Hubble Space Telescope (HST)} imaging \citep[e.g.,][]{Dolphin2002, Weisz2011}. Such imaging was obtained via the HST/ACS Nearby Galaxy Survey Treasury \citep[ANGST;][]{Dalcanton2009}, which surveyed a volume limited sample of $\sim70$ nearby galaxies from 1-4~Mpc in optical ACS and WFPC2 filters. ANGST galaxies show a wide diversity of SFHs \citep[see e.g.,][]{Weisz2011, Williams2013} and present a multitude of candidate TP-AGB stars at a range of metallicities.

\citet[][hereafter G10]{Girardi2010} used ANGST galaxies that showed little-to-no recent SF to constrain lifetimes of low mass, low metallicity TP-AGB stars by suggesting a phase of mass loss before the onset of dust driven winds (pre-dust). \citet[][hereafter R14]{Rosenfield2014} extended the \citetalias{Girardi2010} galaxy sample by incorporating near-infrared imaging from a followup HST/SNAP campaign \citep[AGB-SNAP;][]{Dalcanton2012} and presented a mass loss prescription that best represents the optical and near-infrared (NIR) luminosity functions (LFs) of seven galaxies which also showed little-to-no recent SF.  Both studies used the most likely SF history of each galaxy to produce stellar population synthesis models, which were used to compare the mean predicted TP-AGB lifetimes to the data using the ratio of TP-AGB stars to red giant branch (RGB) stars (\narratio). \citetalias{Rosenfield2014} further performed a calibration of the TP-AGB mass-loss prescriptions by comparing the shapes of the predicted and observed luminosity functions. However, due to the lack of recent SF in the selected galaxies in both programs, the TP-AGB stars studied were restricted to initial masses $\sim0.8-2.5\msun$ and metallicities $-1.54<\feh<-0.86$. In this paper, we extend the analysis of \citetalias{Girardi2010} and \citetalias{Rosenfield2014} to include 10 more galaxies of the AGB-SNAP sample that show evidence of recent SF, and probe a metallicity range of $-1.59 \lesssim \feh \lesssim -0.56$, and an initial mass range of $\sim0.8-4\msun$ of stars that may eventually populate the TP-AGB, importantly including TP-AGB stars that may undergo HBB. The paper is outlined as follows: in Section \ref{sec_data} we briefly describe the observations, photometry, and data reduction used in this project, and present the re-derivation of SFHs and a new method of identifying TP-AGB stars in galaxies with recent SF. In Section \ref{sec_models} we review the \citetalias{Rosenfield2014} TP-AGB model that we test against the observations. We model the data using population synthesis described in Section \ref{sec_model_data}.  We compare the population synthesis models to the observations in Section \ref{sec_analysis}, and conclude in Section \ref{sec_conc}.

\section{Data}
\label{sec_data}
\subsection{Galaxy Sample}

The galaxy sample derives from \citet[][hereafter M12]{Melbourne2012}, who developed 23 resolved stellar source catalogs of 22 galaxies that matched the AGB-SNAP (HST/WFC3) NIR data with the ANGST (HST/ACS or WFPC2) optical data.  We selected 12 fields of 10 galaxies from \citetalias{Melbourne2012} that show evidence of recent SF (and were not one of the 7 galaxies studied in \citetalias{Rosenfield2014}).

Detailed information on the galaxies in this study can be found in  \citet{Dalcanton2009, Dalcanton2012}. Briefly, they are all dwarf spheroidal or dwarf irregular galaxies with the exceptions of NGC~2403 and NGC~300. In the case of NGC~2403, we study a deep (WFPC2) field in the disk of the galaxy as well as a shallower halo (ACS) field \citep[c.f. Figure 2,][]{Williams2013} and we use two fields of UGC~4305 (HoII) with similar SFH. Some parameters of interest of these galaxies are presented in Table~\ref{tab_basic} below, and  typical optical-NIR CMDs from the \citetalias{Melbourne2012} matched catalogs are presented in Figure~\ref{fig:rheb_contam}.

We make use of the ANGST optical data, AGB-SNAP NIR data, the \citetalias{Melbourne2012} matched catalogs and the corresponding artificial star tests (ASTs) in this study. Following the methodology of \citetalias{Rosenfield2014}, we re-calculate the SFHs using the same stellar evolutionary models that we use in our population synthesis code. In order to have best leverage on the SFH, we use the optical ANGST data which reaches several magnitudes below the tip of the RGB (TRGB), usually beyond the red clump. The AGB-SNAP NIR data (as part of the \citetalias{Melbourne2012} matched catalogs) have a more complete census of TP-AGB stars than the optical ANGST data so we use them to compare against our simulated stellar populations. Finally, artificial star tests are employed in the SFH derivation, applied to make completeness corrections to the observed LFs (see Figure \ref{fig:lfs}), and used to simulate statistically similar uncertainties and crowding in the modeled stellar populations. We now briefly discuss the reduction and photometry of each catalog, and the construction of the \citetalias{Melbourne2012} four-filter catalogs.

\subsection{Reduction and Photometry}
The reduction and photometry of the ANGST and AGB-SNAP data as well as the AST strategy are fully described in \citet{Dalcanton2009, Dalcanton2012}. In short, all data were photometered and reduced as part of the University of Washington photometry pipeline, which uses the DOLPHOT package\footnote{\url{http://purcell.as.arizona.edu/dolphot/}} \citep{Dolphin2000} and the ACS, WFC3, or WFPC2 module.  Briefly we use the ``good'' ({\tt gst}) star catalogs  that apply conservative limits of the DOLPHOT flags: {\tt sharpness}, $({\tt sharp_1 + sharp_2})^2 \le 0.075$; {\tt crowding}, ${\tt crowd_1 + crowd_2} \le 0.1$; signal-to-noise, ${\tt S/N} > 4$; and ${\tt flag} < 8$ (i.e., not extended, elongated, highly saturated, or significantly cut off by CCD chip edges).

We refer the reader to \citetalias{Melbourne2012} for complete details on the construction of the four-filter NIR-optical matched catalog and only outline the methodology here. To create the matched catalogs, \citetalias{Melbourne2012} transformed the NIR data set coordinate system to that of the optical using 150 bright, spatially dispersed red stars in the DOLPHOT output catalogs (which were also input to the ANGST {\tt gst} catalogs). The transformation was calculated iteratively (with the {\it MATCH} routine developed by Michael Richmond); first they found a linear fit between the coordinate systems and then used the fit as starting point for a quadratic solution. \citetalias{Melbourne2012} reports 90\% of the stars in the NIR catalogs are matched to within 0.07\arcsec\ of a star in the optical catalogs.

\subsection{Star Formation Histories}
\label{sec_sfh}

Star formation histories were calculated using the {\tt calcsfh} routine in the {\tt MATCH} package \citep{Dolphin2002}. {\tt MATCH} is a CMD-fitting package that finds the most likely SFH and metallicity evolution of an input CMD given an IMF, binary fraction, uncertainties and completeness of the photometry, and stellar evolution models. We adopted the \citet{Kroupa2001} IMF and a binary fraction of 0.35 with a flat distribution of secondary masses. The faint limits of the photometry were set to the 50\% completeness limit determined by the artificial star tests, listed in Table 5 of \citet{Dalcanton2009}. For the stellar evolution models, we use a subset of the  {\bf PA}dova T{\bf R}ieste {\bf S}tellar {\bf E}volution {\bf C}ode (PARSEC V1.2S) stellar evolution tracks \citep{Bressan2012, Bressan2013} (discussed below). Finally, we adopted a binning scheme in color-magnitude space of 0.05 and 0.1 respectively, and metallicity (as $\log$ Z) of 0.15 dex. We further allowed {\tt MATCH} to find the best fitting distance and extinction. However, in the cases where {\tt calcsfh} did not converge to within 0.05 dex of the distance modulus or $A_V$ value derived from TRGB fitting \citep{Dalcanton2009}, we adopted their values to ensure the model CMDs agreed to within a pixel of the binned data.

PARSEC was also used in \citetalias{Rosenfield2014} to derive the SFHs since it calculates the first thermal pulse used in the COLIBRI TP-AGB models \citep{Marigo2013}. Since \citetalias{Rosenfield2014}, PARSEC has undergone significant updates, the largest of which extended the upper mass limit of the models from 12 \msun\ to 350 \msun\ in certain metallicities \citep{Tang2014} and presented new low-mass stellar evolution models \citep{Chen2015}. To import PARSEC for use in {\tt MATCH}, we redefined PARSEC equivalent stellar evolutionary points (i.e., interpolation points along a stellar evolution track) such that each track has a uniform number of points and equidistant spacing between them.  As {\tt MATCH} requires a complete and regular grid in mass and metallicity, this resulted in PARSEC V1.2S in {\tt MATCH} having a different parameter range than the full PARSEC V1.2S. In {\tt MATCH}, the PARSEC V1.2S grid covers a mass range of $0.1\ \msun \le M \le 120\ \msun$ which translates to an age range of $3.55\  \rm{Myr} \le\rm{Age}\le 15.8\ \rm{Gyr}$ and a metallicity range from $-2.18 \le \feh \le 0.42$ (the full grid of PARSEC V1.2S tracks reaches $\feh=0.6$ and younger ages). Finally, the BC transformations of the stellar models occur within {\tt MATCH}, and use those specified in \citet{Girardi2008}, which largely rely on the \citet{Castelli2003} model atmospheres.

With the above parameter search space, we re-fit the program galaxies using the optical data from ANGST and ASTs described in \citet{Dalcanton2009} and  \citet{Weisz2011}. Following the method of \citetalias{Girardi2010} and \citetalias{Rosenfield2014}, we excluded stars above the TRGB from the fitting \citep[using the TRGB magnitudes calculated in][]{Dalcanton2012}. While \citetalias{Girardi2010} found very small or no differences in measured SFH including or not including TP-AGB stars, we excluded them from the data as we had not yet ported model TP-AGB stars into {\tt MATCH} and did not want to introduce bias due to the fitting algorithm finding high S/N data where there were no model points.

\begin{figure*}
\includegraphics[width=\textwidth]{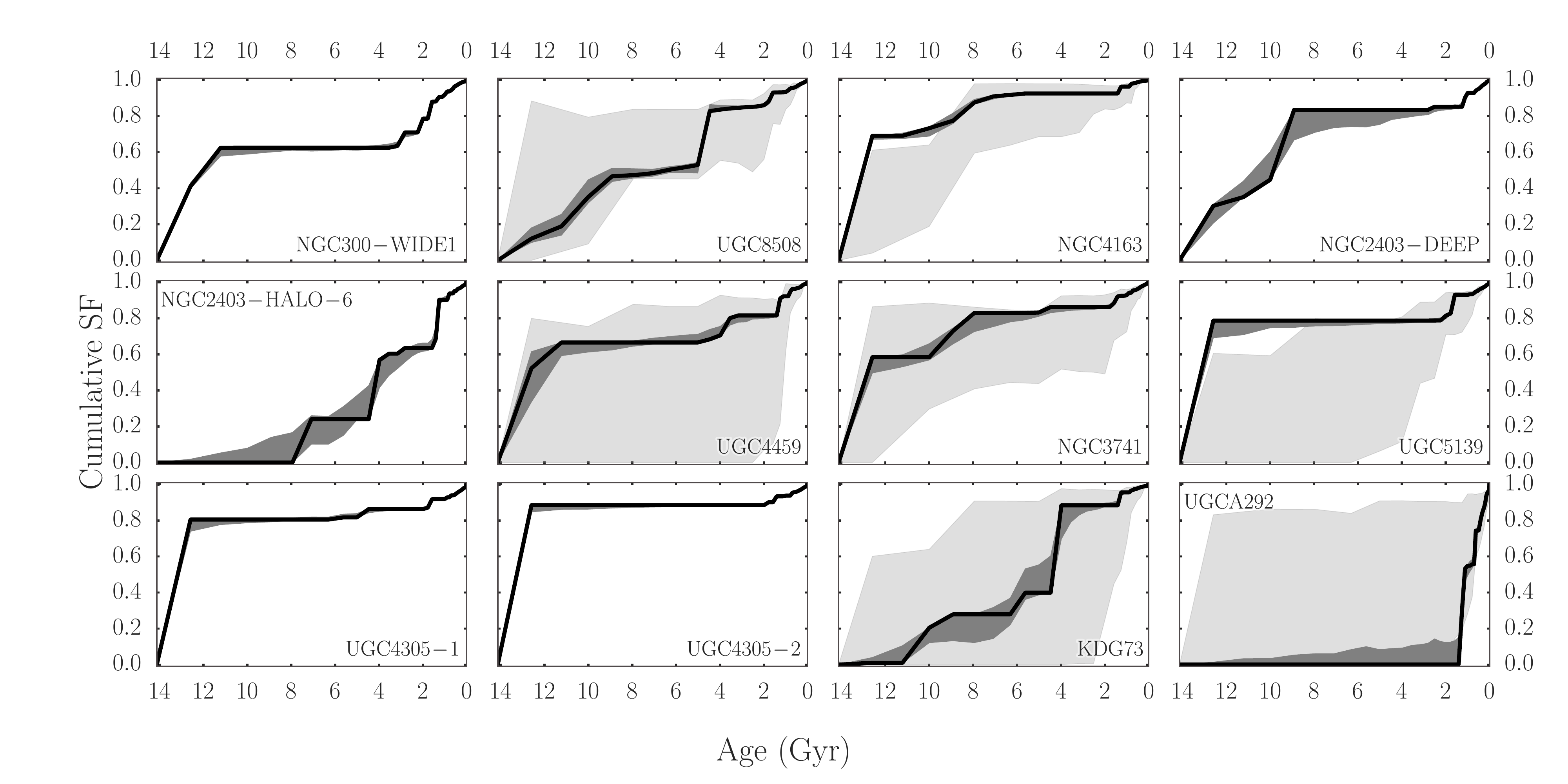}
\caption{{\tt MATCH} solutions for the cumulative SF rate as a function of galaxy age for the galaxies in the sample (black). Dark grey regions denote random uncertainties calculated by {\tt MATCH}'s Hybrid Monte Carlo algorithm \citep[see][]{Dolphin2013}, light gray regions are the systematic uncertainties calculated in \citet[][where available]{Weisz2011}. SFH solutions using PARSEC agree to within systematic uncertainties except for NGC~4163 and UGC~5130, where PARSEC predicts higher SF in the oldest, most uncertain age bins (see text).}
\label{fig_csfr}
\end{figure*}

The last four columns of Table \ref{tab_basic} list the relative amounts of recent SF and corresponding average metallicities for the galaxy sample. Cumulative SFHs are shown in Figure \ref{fig_csfr} with random uncertainties calculated with the Hybrid Monte Carlo algorithm implemented as the {\tt hybridMC} routine in {\tt MATCH} and described in \citet{Dolphin2013}. All SFH fits are consistent to less than 0.01\% level except for the KDG~73 fit which is consistent to 0.12\%.

The program targets already have measured SFH using {\tt MATCH} \citep[][]{Weisz2011, Dalcanton2012, Melbourne2012} with the exceptions of the fields of NGC~2403 and NGC~300 which were separately studied in detail in \citet{Williams2013} and \citet{Gogarten2010}. All of the above previous SFH determinations were calculated with slightly different stellar models \citep[Padua;][]{Girardi2010, Marigo2008}. Therefore it is worth comparing the published results to our new SFH derivations.

To compare the SFHs calculated with either the PARSEC or the Padua models, we overplot the systematic uncertainties from \citet{Weisz2011} in Figure \ref{fig_csfr}. The systematic uncertainties were estimated by deriving the SFH of a synthetic CMD with a different set of stellar evolution models than were inputted and applying the uncertainties to the galaxy's SFH solution. For example, \citet{Weisz2011} produce a synthetic CMD using BaSTI models \citep{Pietrinferni2004} and then derive the SFH of that synthetic CMD using Padua models. The offsets between the input SFH that created the synthetic CMD and the output SFH are related to systematic uncertainties in the stellar models \citep[for full details of this method, see][]{Dolphin2012}. We include systematic uncertainties as a visual guide to the range of possible SFHs, though we do not expect differences from Padua to PARSEC to be nearly as great as are seen between BaSTI-Padua. Furthermore, it should be noted that for these data, using one stellar evolution model instead of another will not produce totally incomparable SFHs \citep[assuming the age, mass and metallicity ranges of the models are comparable, c.f.][]{Dolphin2012}.

The cumulative SFHs are within uncertainties with the exceptions of NGC~4163 and UGC~5139, where \citet{Weisz2011} calculate relatively suppressed SF at ages older than $\sim 10$ Gyr compared to our solutions. The discrepancies between PARSEC and Padua SFH solutions for these two galaxies clearly go beyond the systematic uncertainties calculated by \citet{Weisz2011} at the oldest ages \citep[and therefore lowest masses, where some of the largest upgrades to the Padua models occurred,][]{Chen2014}. However, as we now show, the apparently large deviations at old ages do not affect the relative mass distribution of the TP-AGB and can be neglected for this study.

\begin{figure*}
\includegraphics[width=\textwidth]{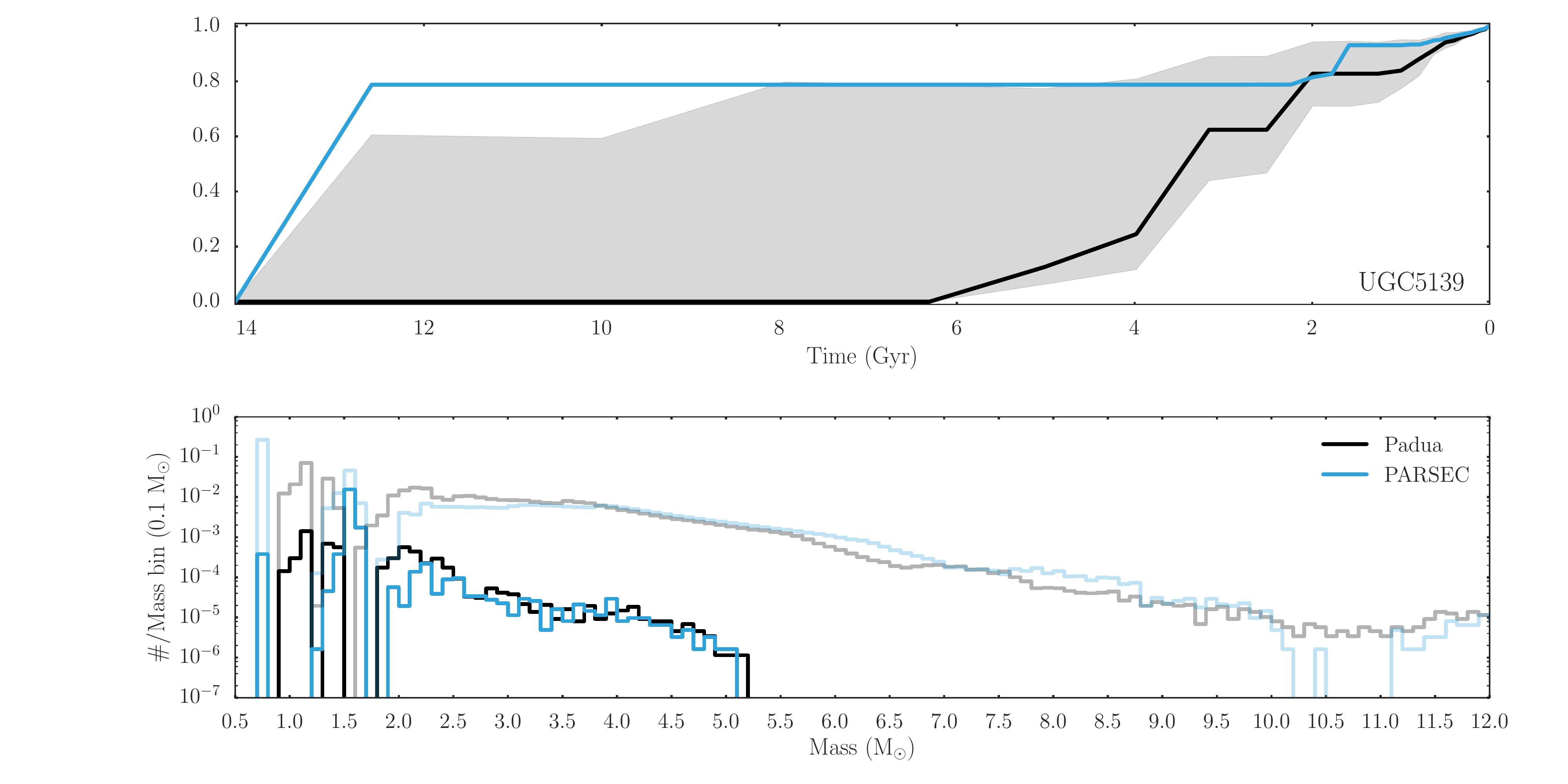}
\caption{Different SFHs due to slight differences in the underlying stellar evolution models do not significantly effect the TP-AGB mass distribution. Top: {\tt MATCH}-derived cumulative SF using optical data of UGC~5139 and stellar evolution models from Padua \citep[black; derived in][]{Weisz2011} and PARSEC (blue; see Figure \ref{fig_csfr}). Significant inconsistencies \citep[outside the bounds of the gray systematic uncertainties derived in][]{Weisz2011} are present at the oldest ages. Bottom: Derived mass distributions using each SFH in the {\tt TRILEGAL} code (see Section \ref{sec_model_data}). The Padua mass distribution is shown in gray with its TP-AGB in black, and the PARSEC mass distribution is shown in light blue with its TP-AGB in blue. Despite the major inconsistencies in the inferred mass distributions, they do not significantly affect the shape of the TP-AGB mass distribution.}
\label{fig_tritest}
\end{figure*}

The top panel of Figure \ref{fig_tritest} shows the cumulative SFH of UGC~5139 reproduced from Figure \ref{fig_csfr} which used PARSEC models as input. Over-plotted is the cumulative SFH from \citet{Weisz2011}, who used Padua models as the stellar evolution models in {\tt MATCH}. In the bottom panel, the inferred mass distributions calculated using {\tt TRILEGAL} and the \citetalias{Rosenfield2014} TP-AGB models (see Section \ref{sec_model_data}) are shown for each SFH in 0.1 \msun\ bins. Despite the clear offsets in the ancient SFR and mass distribution of the inferred stellar population, the relative TP-AGB mass distribution hardly changes. Furthermore, a Kolmogorov-Smirnov test \citep{Kolmogorov1933, Smirnov1948}\footnote{implemented with {\tt ks\_2samp} in {\tt scipy} \citep{scipy}} of the TP-AGB mass distribution resulted in $D_{KS}$ statistic of 0.036 and a $p$-value of 0.9995, therefore we can reject the null hypothesis, which indicates that the SFHs from previous studies using {\tt MATCH}/Padua give statistically indistinguishable TP-AGB populations compared to those produced with {\tt MATCH}/PARSECv1.2s.

\begin{landscape}
\begin{deluxetable}{lcccccccc}
\tablecaption{Basic Parameters of The Galaxy Sample\label{tab_basic}}
\tablehead{
	\colhead{Galaxy} &
	\colhead{Optical Filters} &
	\colhead{A$_V$} &
	\colhead{$(m\!-\!M)_0$} &
    \colhead{$D$} &
	\colhead{$\frac{\rm{SF}}{\rm{SF_{TOT}}}$} &
	\colhead{$\langle \feh \rangle$} &
	\colhead{$\frac{\rm{SF}}{\rm{SF_{TOT}}}$} &
	\colhead{$\langle \feh \rangle$}  \\
	\colhead{} &
	\colhead{} &
	\colhead{} &
	\colhead{} &
    \colhead{(Mpc)} &
	\multicolumn{2}{c}{$<1\rm{Gyr}$} &
	\multicolumn{2}{c}{$1-3\rm{Gyr}$}
}
\startdata
NGC~300-WIDE1 & F606W,F814W &  0.01 &  26.50 &  2.00 & $ 0.08^{+ 0.01}_{- 0.01}$ & $-0.19^{+ 0.12}_{- 0.12}$ & $ 0.29^{+ 0.02}_{- 0.04}$ & $-0.21^{+ 0.12}_{- 0.12}$ \\
UGC~8508 & F475W,F814W &  0.05 &  27.05 &  2.57 & $ 0.07^{+ 0.01}_{- 0.01}$ & $-1.19^{+ 0.12}_{- 0.12}$ & $ 0.09^{+ 0.03}_{- 0.02}$ & $-1.22^{+ 0.12}_{- 0.12}$ \\
NGC~4163 & F606W,F814W &  0.06 &  27.29 &  2.87 & $ 0.04^{+ 0.00}_{- 0.00}$ & $-0.99^{+ 0.12}_{- 0.12}$ & $ 0.04^{+ 0.00}_{- 0.00}$ & $-1.00^{+ 0.12}_{- 0.12}$ \\
NGC~2403-DEEP & F606W,F814W* &  0.12 &  27.50 &  3.16 & $ 0.07^{+ 0.01}_{- 0.01}$ & $-0.49^{+ 0.12}_{- 0.12}$ & $ 0.09^{+ 0.02}_{- 0.04}$ & $-0.53^{+ 0.12}_{- 0.12}$ \\
NGC~2403-HALO-6 & F606W,F814W &  0.12 &  27.50 &  3.16 & $ 0.10^{+ 0.02}_{- 0.03}$ & $-0.78^{+ 0.13}_{- 0.12}$ & $ 0.30^{+ 0.07}_{- 0.08}$ & $-0.81^{+ 0.16}_{- 0.13}$ \\
UGC~4459 & F555W,F814W &  0.12 &  27.79 &  3.61 & $ 0.08^{+ 0.01}_{- 0.01}$ & $-0.99^{+ 0.12}_{- 0.35}$ & $ 0.11^{+ 0.03}_{- 0.01}$ & $-1.02^{+ 0.12}_{- 0.12}$ \\
NGC~3741 & F475W,F814W &  0.08 &  27.55 &  3.24 & $ 0.07^{+ 0.01}_{- 0.01}$ & $-1.19^{+ 0.12}_{- 0.12}$ & $ 0.07^{+ 0.02}_{- 0.02}$ & $-1.20^{+ 0.12}_{- 0.12}$ \\
UGC~5139 & F555W,F814W &  0.15 &  27.95 &  3.89 & $ 0.07^{+ 0.01}_{- 0.01}$ & $-0.80^{+ 0.12}_{- 0.12}$ & $ 0.14^{+ 0.02}_{- 0.03}$ & $-0.89^{+ 0.12}_{- 0.12}$ \\
UGC~4305-1 & F555W,F814W &  0.01 &  27.65 &  3.39 & $ 0.08^{+ 0.01}_{- 0.01}$ & $-1.18^{+ 0.12}_{- 0.12}$ & $ 0.06^{+ 0.01}_{- 0.01}$ & $-1.18^{+ 0.12}_{- 0.12}$ \\
UGC~4305-2 & F555W,F814W &  0.10 &  27.70 &  3.47 & $ 0.06^{+ 0.00}_{- 0.00}$ & $-1.08^{+ 0.12}_{- 0.12}$ & $ 0.05^{+ 0.01}_{- 0.01}$ & $-1.09^{+ 0.12}_{- 0.12}$ \\
KDG~73 & F475W,F814W &  0.06 &  28.03 &  4.04 & $ 0.04^{+ 0.01}_{- 0.01}$ & $-1.20^{+ 0.12}_{- 0.12}$ & $ 0.07^{+ 0.04}_{- 0.04}$ & $-1.28^{+ 0.12}_{- 0.12}$ \\
UGCA~292 & F606W,F814W &  0.00 &  28.64 &  5.35 & $ 0.45^{+ 0.06}_{- 0.05}$ & $-1.20^{+ 0.12}_{- 0.12}$ & $ 0.55^{+ 0.07}_{- 0.23}$ & $-1.27^{+ 0.14}_{- 0.12}$ \\
\enddata
\tablecomments{Properties of the Galaxy Sample. The first column lists the target consistent with the footprints named in \citet{Dalcanton2009, Dalcanton2012} and available on the Mikulski Archive for Space Telescopes (MAST), followed by the optical filters used in the {\tt MATCH} CMD-fitting ($^*$WFPC2 data). Columns 3 and 4 are the extinction, $A_V$, and distance modulus, $(m-M)_0$, \citep[calculated from TRGB fitting,][]{Dalcanton2009}. Columns 5 and 7 show the fraction of total stars formed between 3.5 Myr and 1 Gyr, and between 1 Gyr and 3 Gyr respectively, Columns 6 and 8 show the average metallicity in the same time bins. (N.B. the youngest PARSEC models currently available in {\tt MATCH} are 3.5 Myr.)}
\end{deluxetable}
\end{landscape}

\subsection{Identifying TP-AGB Stars}
\label{sec_ids}
\begin{figure*}
\includegraphics[width=0.49\textwidth]{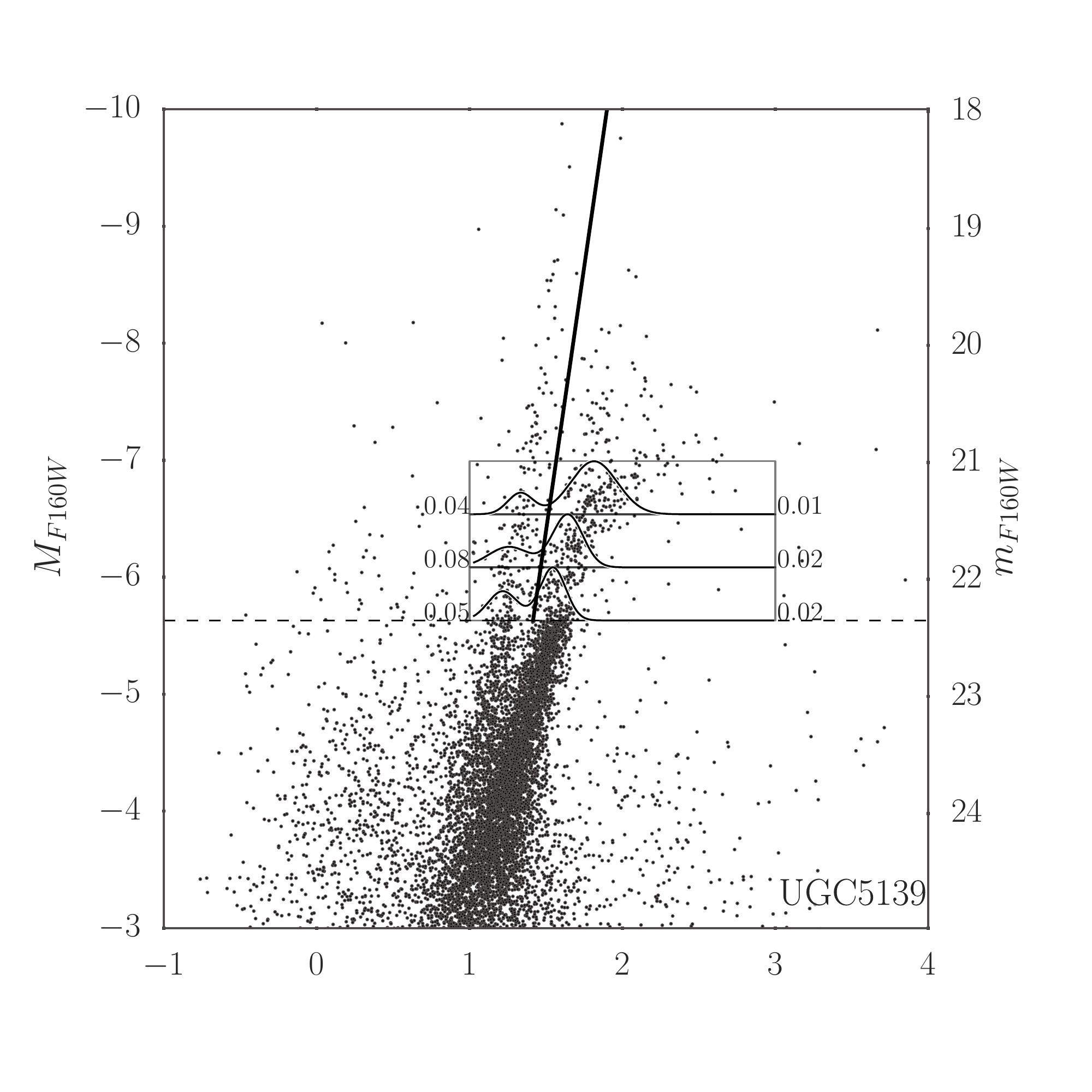}
\includegraphics[width=0.49\textwidth]{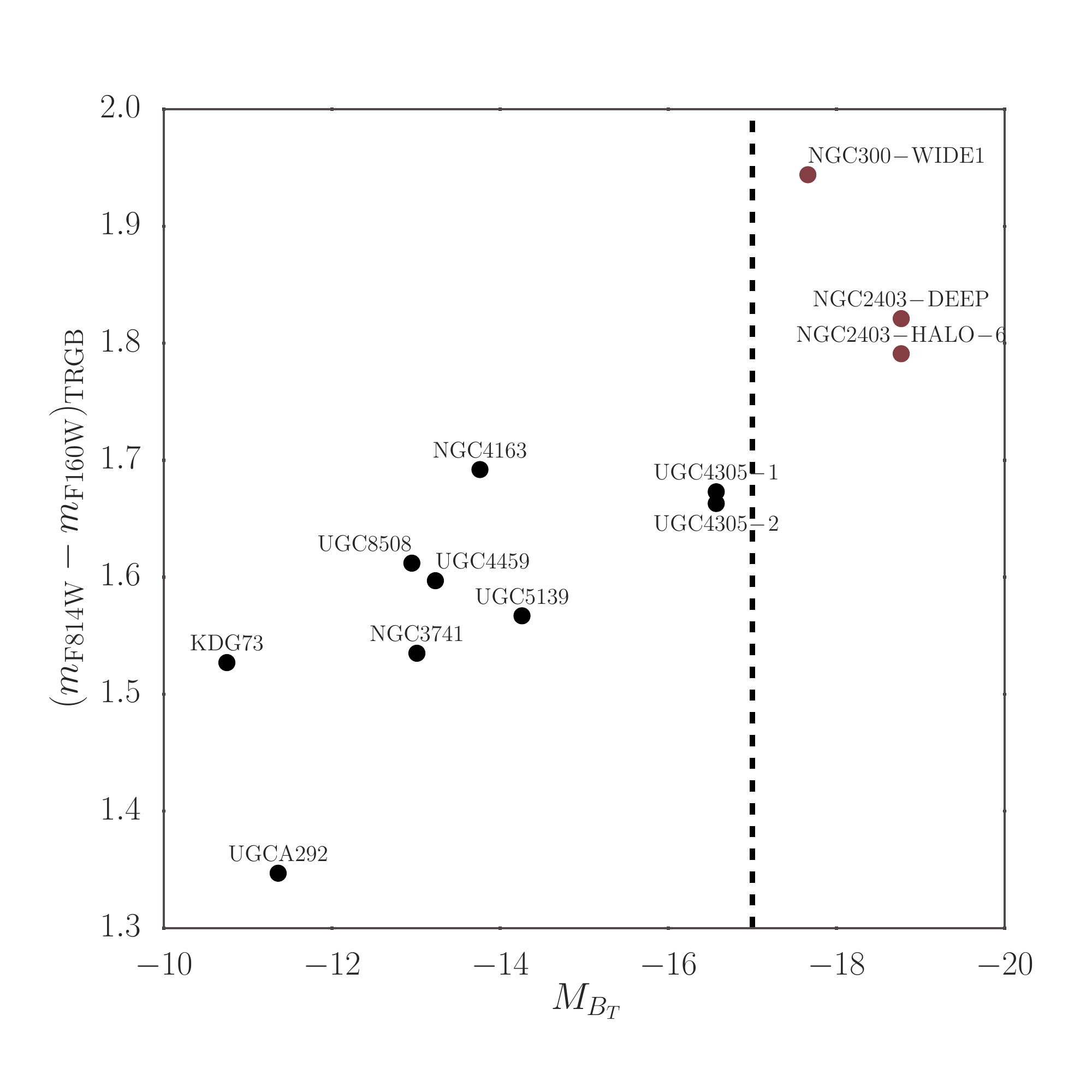} \\
\includegraphics[width=0.48\textwidth]{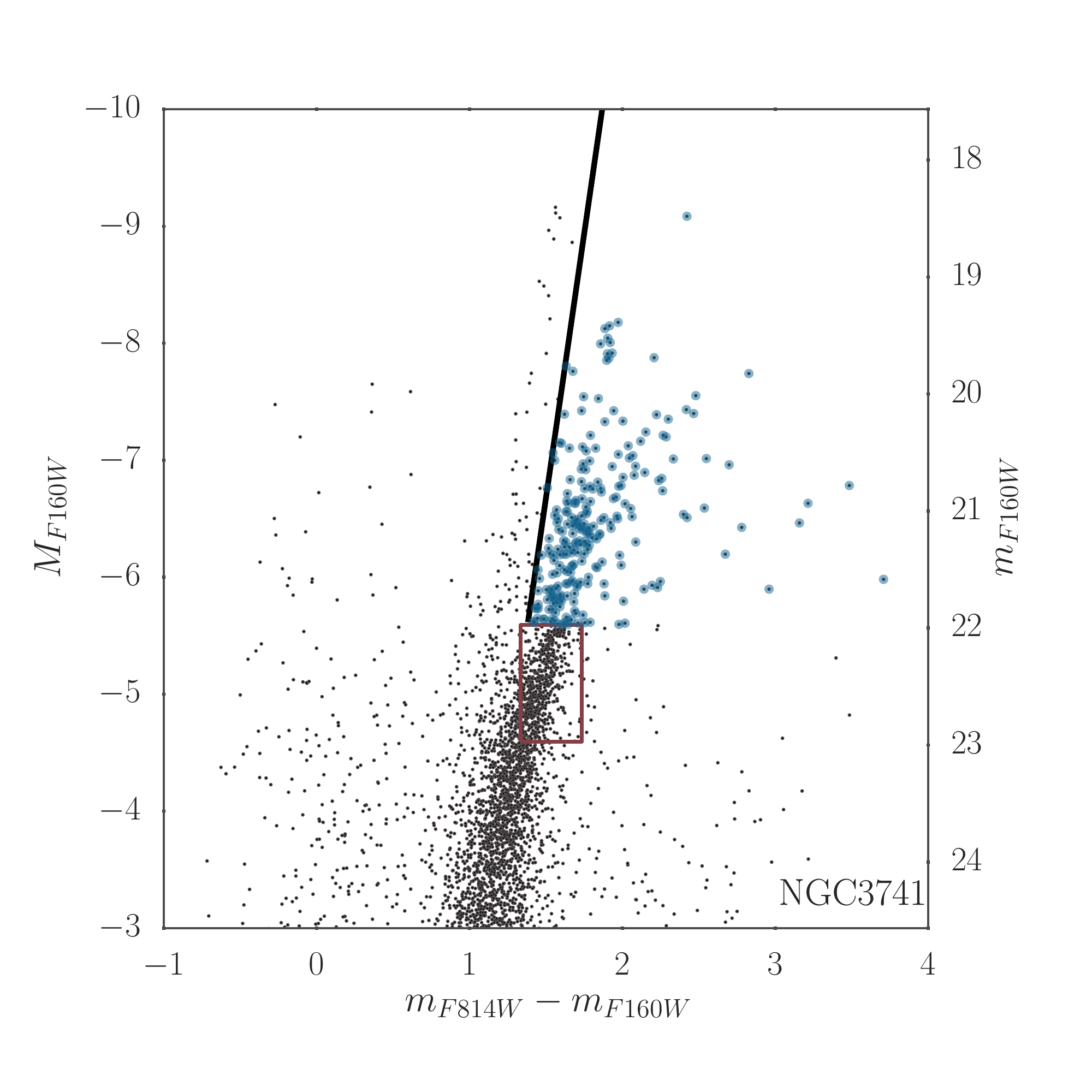}
\includegraphics[width=0.48\textwidth]{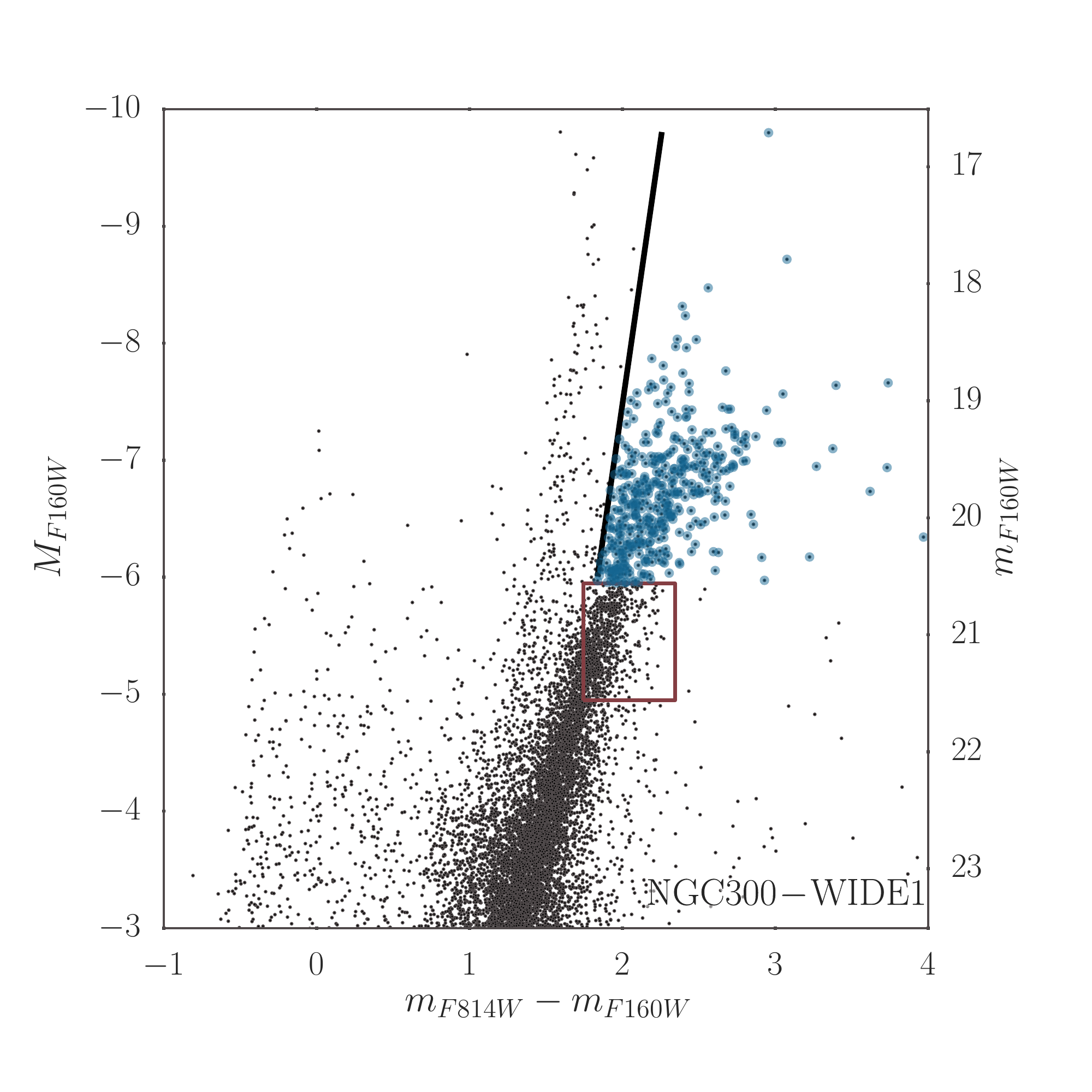}
\caption{Selecting TP-AGB and RGB stars. {\it {Top left}}: Optical-NIR absolute magnitude CMD of UGC~5139 with the results of double Gaussian fitting method discussed in the text (thin black lines) to test the separation line between RHeB and TP-AGB stars (thick black line). Next to each double Gaussian are two sets of numbers in each magnitude bin, on the left is the fraction of stars from the ``RHeB'' Gaussian that are expected in the TP-AGB region, and the right is the fraction of stars in the ``TP-AGB'' Gaussian expected in the RHeB region. The TRGB magnitude is marked by a dashed horizontal line. {\it Top right}: Optical-NIR TRGB color vs $M_{B_T}$, galaxies brighter than $M_B=-17$ have a wider color selection box for the RGB than those fainter. {\it Bottom}: Example results of the TP-AGB and RGB identification for NGC~3741 (left) and NGC~300 (right). RGB stars selected to scale the population synthesis models are within the red box and stars considered TP-AGB (above the TRGB and red-ward of the RHeB-TP-AGB line) are shown in blue, with the RHeB-TP-AGB line shown in black.\label{fig:rheb_contam}}
\end{figure*}

The region of TP-AGB stars in an optical-NIR CMD are easy to identify by eye, as TP-AGB stars are the brightest and reddest when present (see Figure \ref{fig:rheb_contam}). However the boundary in CMD space near the TRGB as well as the blue edge of TP-AGB can contain RGB stars and red core Helium burning (RHeB) stars, respectively. Confusing RGB and RHeB stars as TP-AGB stars will erroneously increase the apparent number of TP-AGB stars in the observations which will be interpreted as longer TP-AGB lifetimes.

We exclude RGB and RHeB stars and thereby isolate TP-AGB stars in the following way. First, to exclude RGB counts in the \narratio\ ratio we continue the method of \citetalias{Rosenfield2014} and exclude a $\pm 0.2\ m_{\rm{F160W}}$ band around the TRGB. To minimize the contribution of RHeB stars counted as TP-AGB we used the \citetalias{Melbourne2012} matched catalogs to find an approximate separation line in CMD space between TP-AGB stars and the RHeB sequence.

We first estimated the RHeB TP-AGB separation line for all of the galaxies stacked together and then tested its adequacy on each individual galaxy to refine the slope and intercept.

To stack the galaxies, we shifted all the optical-NIR data to absolute magnitudes using the distance modulii and extinction values ($A_V$) calculated in \citet{Dalcanton2009} and \citet{Dalcanton2012} and corrected the $A_V$ values to the extinction values of the filters in question using coefficients derived from \citet{Cardelli1989} with an extinction curve with R$_V$=3.1 applied to a G2V star \citep[$A_{F160W}/A_V=0.6056$ and $A_{F814W}/A_V=0.2044$, see][]{Girardi2008}.

We tested the initial separation line by fitting a double Gaussian distribution as a function of absolute $F160W$ magnitude for each field within the color limits $F814W\!-\!F160W=[1, 3]$ using Levenberg-Marquardt least-squares minimization\footnote{the python-implementation based on {\tt MINPACK-1} \url{http://cars9.uchicago.edu/software/python/mpfit.html}}. We then calculated the expected numbers of stars from each Gaussian on each side of the separating line for each magnitude bin within the color limits.

We set the magnitude bins with the binning algorithm from \citet{Knuth2006} implemented in the {\tt astroML} package \citep{astroML}. The resulting slope works well in many of the most populated systems, however it fails in the regime where a single Gaussian can provide a better fit than a double, a scenario usually due to an under-populated RHeB or a large color dispersion of the TP-AGB (a sign of an intrinsic metallicity spread).

We also found that the slope calculated worked well for all galaxies, but we needed to further take into account the intrinsic offsets in the color of the RGB from galaxy to galaxy due differences in metallicity. Therefore, we adjusted the separation line to include a TRGB color term. The resulting line can be reproduced following the equation:

\begin{eqnarray*}
M_{\rm F160W} = - 7.0 - 9.03 \big[(M_{\rm F814W}\!&-&\!M_{\rm F160W}) - \\
								  (M_{\rm F814W}\!&-&\!M_{\rm F160W})_{\rm{TRGB}}\big] \ .
\end{eqnarray*}

An example of this procedure is shown in Figure \ref{fig:rheb_contam}. With this method, we find for each field, RHeB stars do not contribute to the TP-AGB by more than 5\% and conversely, fewer than 6\% of TP-AGB stars are found in the RHeB region (mean of 3\%). We verified this result using {\tt TRILEGAL} in stellar population synthesis mode and added photometric uncertainties from the ASTs (the same method as discussed below in Section \ref{sec_models}). Since we are able to track the individual stellar phases in the {\tt TRILEGAL} simulated stellar catalog, we find the mean contamination by RHeB stars to be 3\% and the fraction of AGB stars (both TP-AGB and Early-AGB) ``lost'' to the RHeB region is never more than 4\%.

Stars red-ward of the separation line and brighter than the TRGB are considered TP-AGB stars in the following analysis. The number of TP-AGB stars identified in each galaxy field is listed in the first two columns of Table \ref{tab_fracs}.

\section{TP-AGB Evolution Models}
\label{sec_models}
The TP-AGB models tested here are essentially the same as in \citetalias{Rosenfield2014}, the full set spans initial TP-AGB masses from $0.50<M<6.0\ M_\odot$ and a metallicity range of $0.0005 < Z < 0.06$ ($-1.48 < \feh  < 0.60$). Obviously, we are limited by the metallicities in the data set and only test the lower end of the calculated metallicity range. Briefly, the TP-AGB models incorporate the following main characteristics:
\begin{itemize}
\item The conditions at the first thermal pulse on the AGB are extracted from PARSEC V1.2S stellar evolution models.
\item The TP evolution is computed with the COLIBRI code which is extensively described in \citet{Marigo2013}. The code includes several innovative features for instance, detailed nucleosynthesis calculations both during the thermal pulses and the interpulse phase, and on-the-fly computation of atomic and molecular abundances and their Rosseland mean opacities \citep{Marigo2009} as the envelope composition changes along the evolution due to mixing events and/or HBB. A few key processes are parameterized, such as the efficiency of third dredge-up episodes and the efficiency of mass loss.
\item A complete nuclear network that includes the proton-proton chains, the CNO, NeNa, MgAl cycles and several $(\alpha,\gamma)$ and $(\alpha,n)$ reactions is coupled to a time-dependent diffusive description of convection to follow the nucleosynthesis in the HBB phases of massive TP-AGB stars. The concentrations of the most abundant elements in the pulse-driven convection zone (e.g., $^4$He, $^{12}$C, $^{16}$O, $^{22}$Ne, $^{24,25}$Mg) are also predicted solving a proper nuclear network.

\item For the mass loss, we adopt a 2-phase formalism, which includes: the pre-dust phase ($\dot{M}_{pd}$) using a modified version of the \citet{Schroeder2005} formula, inspired by the theoretical work of \citet{Cranmer2011} on Alfv\`en waves and atmospheric turbulence; and the pulsation-assisted dust-driven phase ($\dot{M}_{dd}$) following a formula similar to \citet{Bedijn1988} but with parameters calibrated with a sample of Galactic Miras. The super-wind mass-loss ($\dot{M}_{sw}$) is assumed not to exceed a maximum limit which is determined according to \citet{Vassiliadis1993}. The largest mass loss between these two prescriptions is adopted as the reported mass loss.
\end{itemize}

\begin{figure}
\includegraphics[width=\columnwidth]{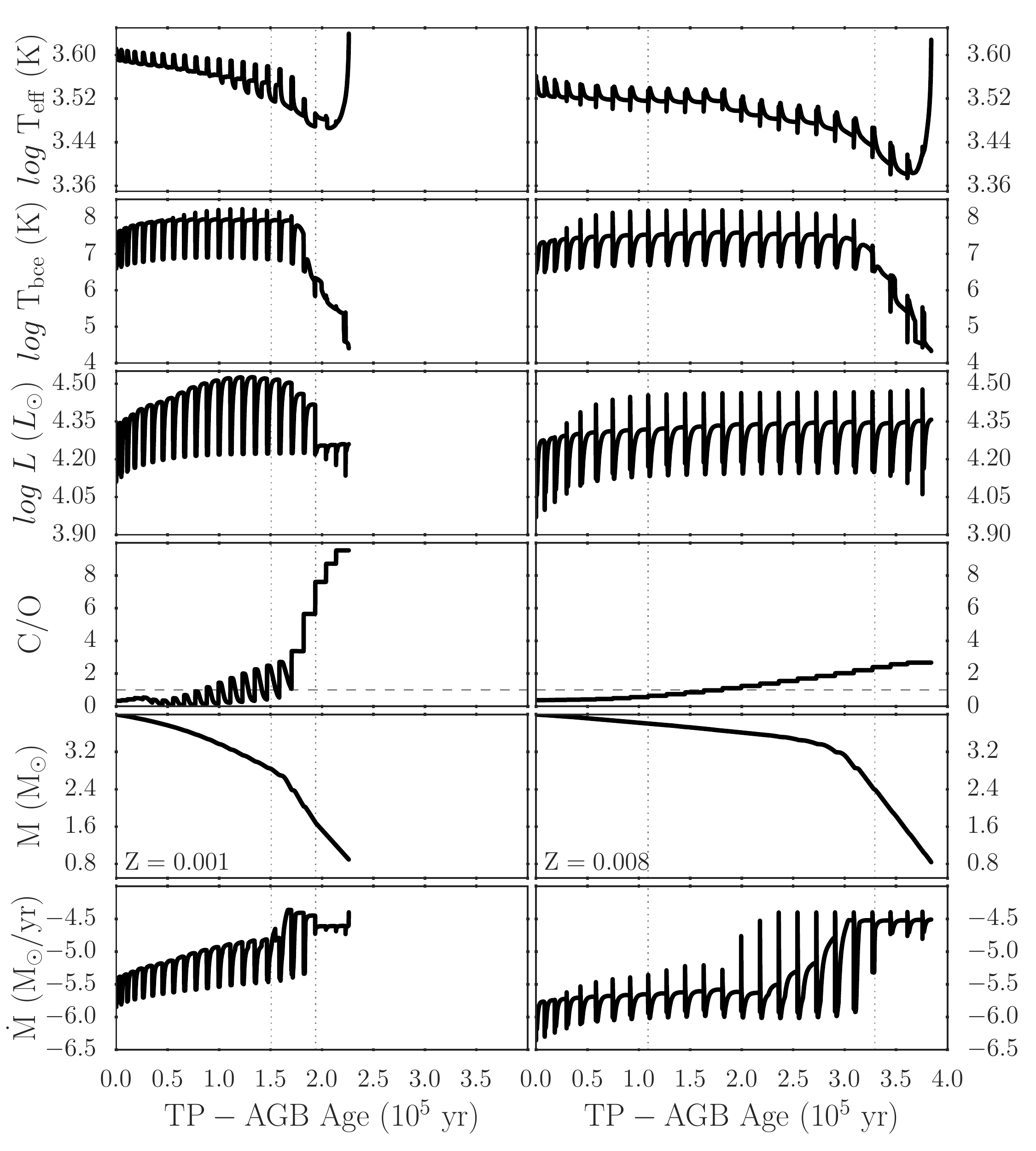}
\caption{Evolution of $4 \msun$ TP-AGB stars at different metallicities, $Z=0.001$ (left) and $Z=0.008$ (right). From top to bottom the panels show effective temperature, temperature at the base of the convective envelope, luminosity, $C/O$ ratio (with $C/O=1$ as a dotted horizontal line), star mass, and mass loss rate. HBB is stronger at lower metallicities due to higher temperatures at the base of the convective zone. HBB does not occur in the higher metallicity TP-AGB star, which exhibits a constant increase in the $C/O$ ratio and does not show the bell-shape increase in stellar luminosity exhibited by the lower metallicity star.}
\label{fig:tpagb_track}
\end{figure}

Low mass TP-AGB stars with the above mass loss prescriptions are fully discussed in \citetalias{Rosenfield2014} so we remark on the higher mass TP-AGB stars. The evolution of two TP-AGB stars with initial mass $M=4.0M_\odot$ and metallicities $Z=0.001$ and $Z=0.008$ ($\feh=-1.18, -0.28$) are shown in Figure \ref{fig:tpagb_track}. Each vertical panel shows a different feature as a function of TP-AGB age \citep[motivated by][]{Vassiliadis1993}, from top to bottom: the effective temperature, temperature at the base of the convective envelope, luminosity, surface Carbon to Oxygen ratio ($C/O$; with a horizontal dashed line at $C/O=1$), stellar mass, and mass loss rate. Vertical dotted lines separate the mass-loss regimes discussed above. The over luminous structure of the lower metallicity track (second from the top panel, left side of Figure \ref{fig:tpagb_track}) is an effect of strong HBB which increases with decreasing metallicity (for a given stellar mass) due to higher temperatures at the base of the convective envelope.

\begin{figure*}
\includegraphics[width=\textwidth]{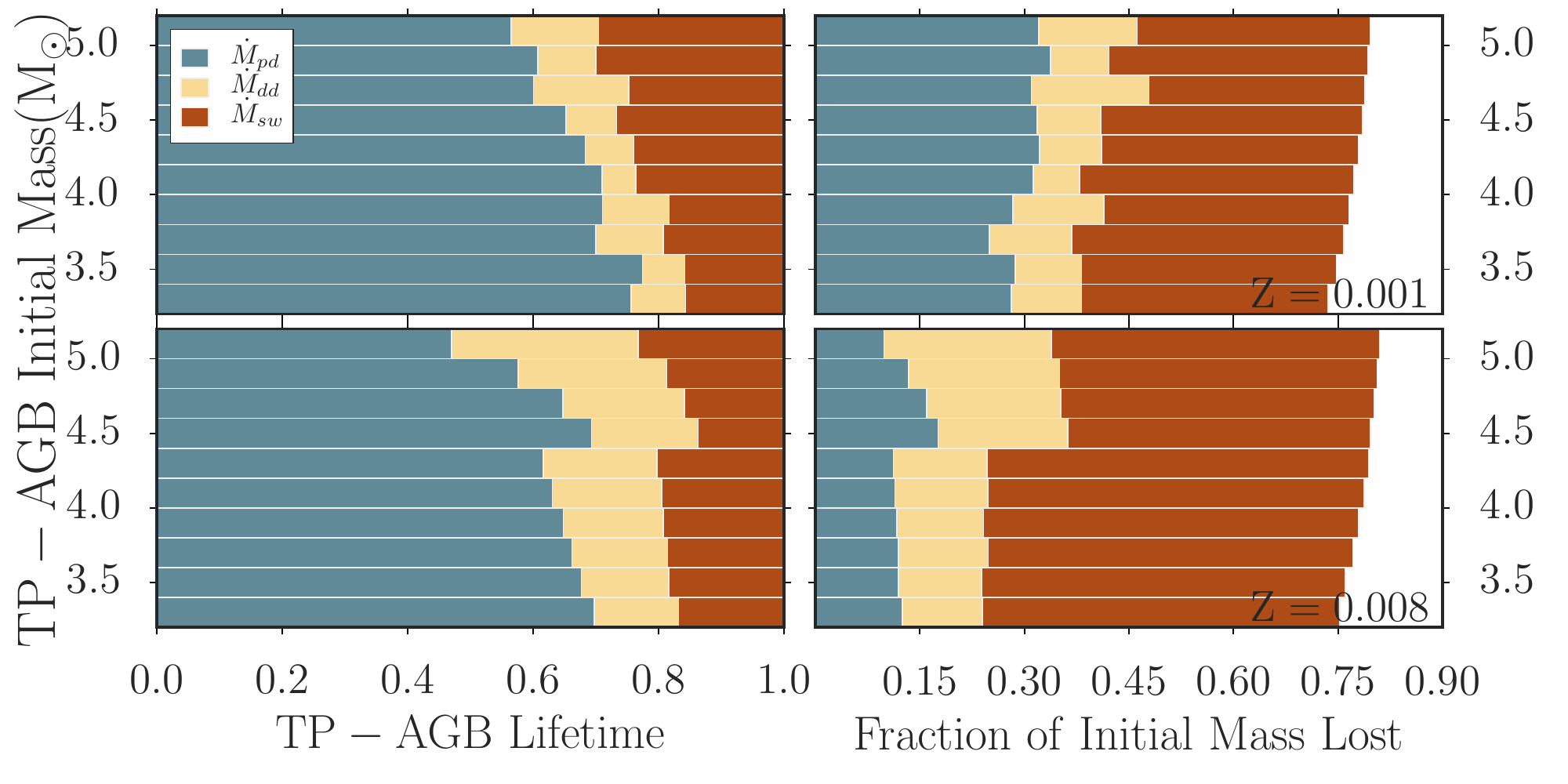}
\caption{Duration of (left) and mass lost in (right) each mass-loss phase for TP-AGB stars (pre-dust, $\dot{M}_{pd}$, in blue; pulsation-assisted dust-driven, $\dot{M}_{dd}$, in yellow; and super-wind, $\dot{M}_{sw}$, in red). Shown are initial masses between $3-5\ \msun$ at different metallicities, $Z=0.001, 0.008$ (top, bottom). Pre-dust mass loss is less efficient than the other phases, however it lasts for the largest proportion of intermediate-mass TP-AGB lifetimes. This trend makes the mass lost to pre-dust winds comparable to the mass lost during the dust-driven winds for lower masses and higher metallcities. For higher masses and lower metallicities, the mass lost in pre-dust winds is comparable to that lost in the super-wind phase.}
\label{fig:duration_masslost}
\end{figure*}

Pre-dust mass loss is important across the full span of TP-AGB mass and metallicities, though decreasingly so with increasing metallicity. Figure \ref{fig:duration_masslost} shows TP-AGB stars with initial masses between $3-5\ \msun$ (calculated at mass steps of $\Delta M=0.2\msun$ at two low metallicities ($Z=0.001, 0.008$ or $\feh=-1.18, -0.28$). The left panels show the fraction of time a TP-AGB star of a given initial mass spends in the pre-dust, pulsation-assisted dust-driven, and super wind phases. The right panels shows the amount of mass lost in each of those phases as a function of initial TP-AGB mass. Generally, the amount of time spent in the pre-dust phase increases with decreasing initial mass while the amount of mass lost from the pre-dust phase is roughly constant with mass (with the exception of the high mass Z=0.008 TP-AGB stars, which shows a slight decrease in mass lost with increasing initial TP-AGB mass). With increasing metallicity, we can expect an overall decrease in the duration of the pre-dust phase as it becomes easier to make dust at lower effective temperatures, so the pulsation-assisted dust-driven wind phase will take over sooner. It is interesting to note the amount of mass lost in the low metallicity pre-dust phase is comparable to the mass lost in the super wind phase, and at increasing metallicity, the pre-dust winds are responsible for a comparable amount of mass lost as is lost to the pulsation-assisted dust driven winds. However, the trends described above are derived with a simplified prescription for AGB mass loss. We postpone to future works the application of state-of-the-art dynamical models of AGB atmospheres \citep{Hoefner2008, Bladh2014, Eriksson2014} for more accurate and physically-grounded predictions of the relative effects of the pre-dust wind and the dust-driven wind as a function of mass an metallicity.

\section{Modeling the Data}
\label{sec_model_data}
To model the data with TP-AGB stars, we use the population synthesis code {\tt TRILEGAL}, initially designed to simulate foreground stars in the Milky Way \citep{Girardi2005}, it is also able to simulate galaxies of known distance, mean extinction and SFH. {\tt TRILEGAL} uses as input the PARSEC V1.2S models. Although such models are computed at constant mass, mass loss during the RGB is taken into account internally by TRILEGAL at the phase of isochrone construction. We adopt RGB mass loss described by \citet{Reimers1975} with a multiplicative factor $\eta=0.2$ \citep{Miglio2012}, which typically produces just a modest (of the order of $\sim0.2$~\Msun\ at most) change in the initial mass of the RGB and TP-AGB stars considered in this work\footnote{As mentioned in passing in \citetalias{Rosenfield2014}, the \citetalias{Rosenfield2014} pre-dust mass loss prescription may be too efficient to extend to the lower luminosities of RGB stars.}.

Since \citet{Girardi2007b}, {\tt TRILEGAL} incorporates TP-AGB stars taking into account a series of their most basic evolutionary and spectral properties. For the present paper, suffice it to mention that the photometric properties are simulated taking fully into account the presence of $L$ and \Teff\ variations due to the thermal-pulse cycles, and the changes both in the photospheric spectra and in dust-reprocessing properties taking place as the stars transit between O-rich and C-rich configurations (with either C/O$<$1 or $>$1, where C/O is the surface carbon-to-oxygen number ratio). A detailed description of how these processes are implemented can be found in \citet{Marigo2008} and \citet{Aringer2009}. In particular, here we assume the dust properties of the 60\% aluminum oxyde + 40\% silicate mixture for O-rich stars, and 85\% amorphous carbon + 15\% silicon carbide mixture for C-rich stars \citep{Groenewegen2006}. In addition, the present code presents some relatively minor updates that are discussed in \citet{Marigo2015}. We ignore the photometric changes due to the long-period variability, since its effect is relatively small in the NIR, and since at low metallicities a major fraction of the TP-AGB stars pulsate just in low-amplitude overtone modes. Therefore, pulsation is not expected to change the bulk numbers nor the LFs of the TP-AGB stars we predict for the present sample of galaxies.

For each galaxy, we follow the same method as \citetalias{Rosenfield2014}, briefly, we sample the random uncertainties in the SFH to produce $\sim100$ {\tt TRILEGAL} stellar populations of an arbitrarily large total mass and apply photometric uncertainty and completeness information from the data by applying ASTs to each simulation. To account for the mass difference between our simulation and the observation, we scale the simulation to match the number of a subset of RGB stars in \citetalias{Melbourne2012}'s four filter matched catalogs.

To select the RGB stars to scale the simulation, unlike \citetalias{Rosenfield2014}, we can not uniformly apply the same $F110W\!-\!F160W$ color cut to exclude possible main sequence contaminants, as the galaxies in our program also have a large number of RHeB stars. Instead, the RGB subset is defined by a box in optical-NIR CMD space with magnitude limits set at the $F160W$ TRGB and 1 mag below. The blue $F814W\!-\!F160W$ limit of the RGB box is set at 0.2 mag bluer than the $F814W-F160W$ TRGB color, which nicely maps to the integrated B-band magnitude, $M_{B_T}$ \citep{Karachentsev2004} as the slope of the RGB is primarily set by the metallicity \citep[e.g, see Section 6 of][]{Dalcanton2012}. Finally, the width of the RGB box is set to 0.4 mag for galaxies with $M_{B_T} > -17$ and 0.6 mag for galaxies brighter (see Figure \ref{fig:rheb_contam}), as it is reasonable to expect at constant T Type, higher mass galaxies had more metallicity evolution, and therefore show a larger spread in RGB color.

\section{Analysis}
\label{sec_analysis}

Using many galaxies to constrain TP-AGB evolution has added benefits beyond statistically large samples of TP-AGB stars. The more galaxies available, the less the resultant constraints are subject to a specific TP-AGB population and a specific SFH. However, these benefits come with their own limitations, we only derive model constraints from the mean properties of the observed TP-AGB populations and therefore do not tune models of one specific combination of age, mass, and metallicity. Our models are deemed satisfactory when they reach broad agreement to the largest span of age, metallicity, and mass accessible from our galaxy sample. Below we present three comparisons of the \citetalias{Rosenfield2014} models to the galaxy sample. First we calculate the relative number of TP-AGB stars to RGB stars, the \narratio~ ratio; second, we compare the shape of the observed and predicted LFs, and finally we compare the flux contribution of TP-AGB stars compared the integrated $F160W$ flux of the image.

\subsection{Ratio of TP-AGB to RGB Stars}

The comparison between the observed and predicted \narratio\ ratio is a rough estimate how well the predicted mean lifetimes of TP-AGB stars agree with those observed. The general assumption is that the number of stars in a given portion of a CMD is related to the amount of time they spend in that portion of the CMD. If by experimental design, the CMD filters are set to optimize the separation of stellar evolution phases, this allows researchers to measure relative lifetimes of different evolutionary phases. Table \ref{tab_fracs} lists the numbers of TP-AGB stars found in the data (column 1) and predicted in the model (column 2), and the \narratio\ for each field (column 3). The model \narratio\ listed for each field is the median of the $\sim100$ {\tt TRILEGAL} models (column 4). The fractional differences between model and data (column 5) serve as a comparison. Each column also contains Poisson uncertainties. A fractional difference of 0 would mean a perfect agreement and negative values indicate a relative overproduction of model TP-AGB stars compared to the number observed.

From field to field, the fractional difference varies largely, from nearly 65\% over production of TP-AGB stars (UGCA~292) to nearly 40\% underproduction (KDG~73). Due to uncertainties in SFH (see Section \ref{sec_sfh}) and variations in TP-AGB populations from galaxy to galaxy, it is more useful to combine the measurements for a rough estimate on the adequacy of the model. Taking the mean \narratio\ of the data and model shows a fractional difference of 9\%, meaning the \citetalias{Rosenfield2014} models show good agreement though slightly under predict the mean TP-AGB lifetimes based on these observations.

\subsection{Luminosity Functions in the Optical and NIR}
\begin{figure*}

\includegraphics[width=0.49\textwidth]{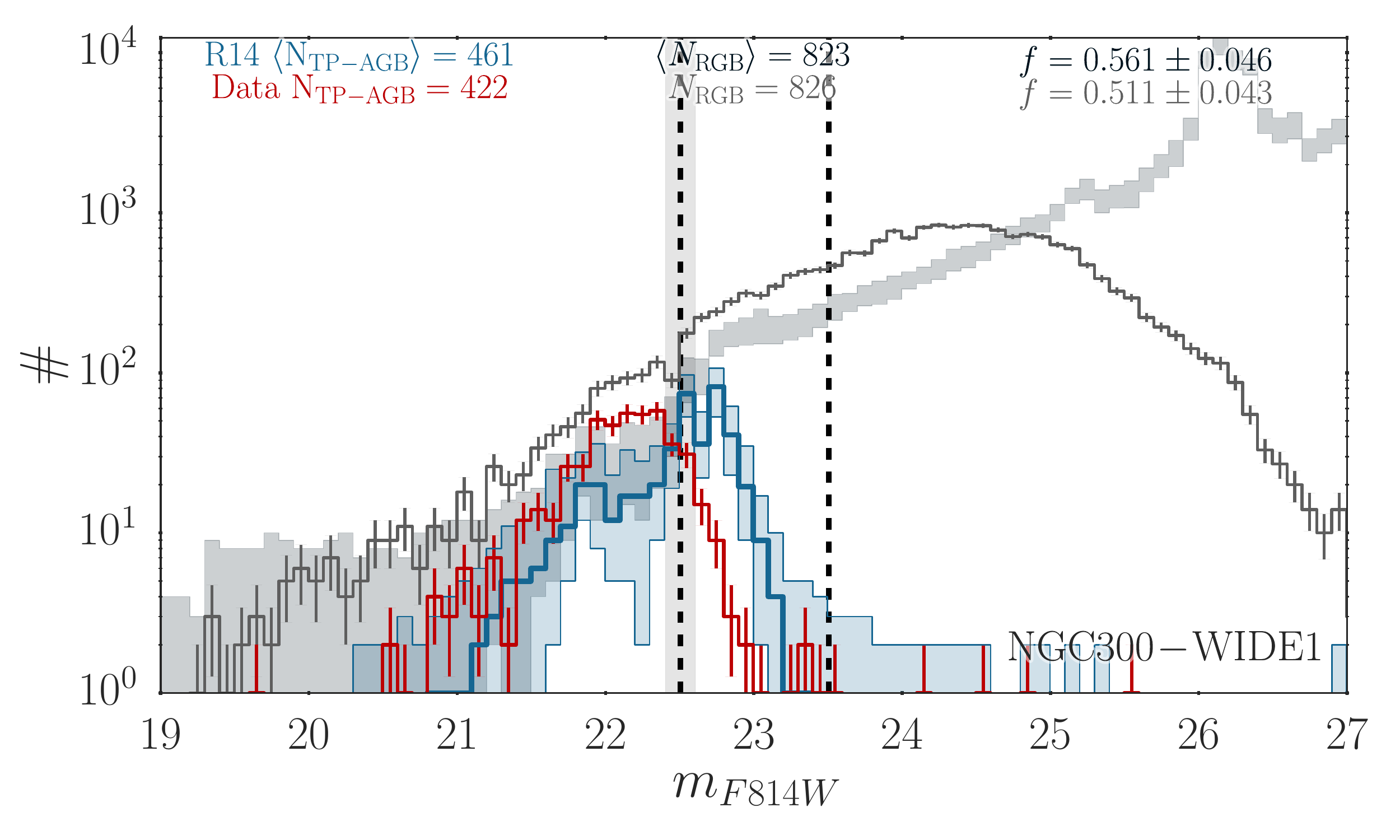}
\includegraphics[width=0.49\textwidth]{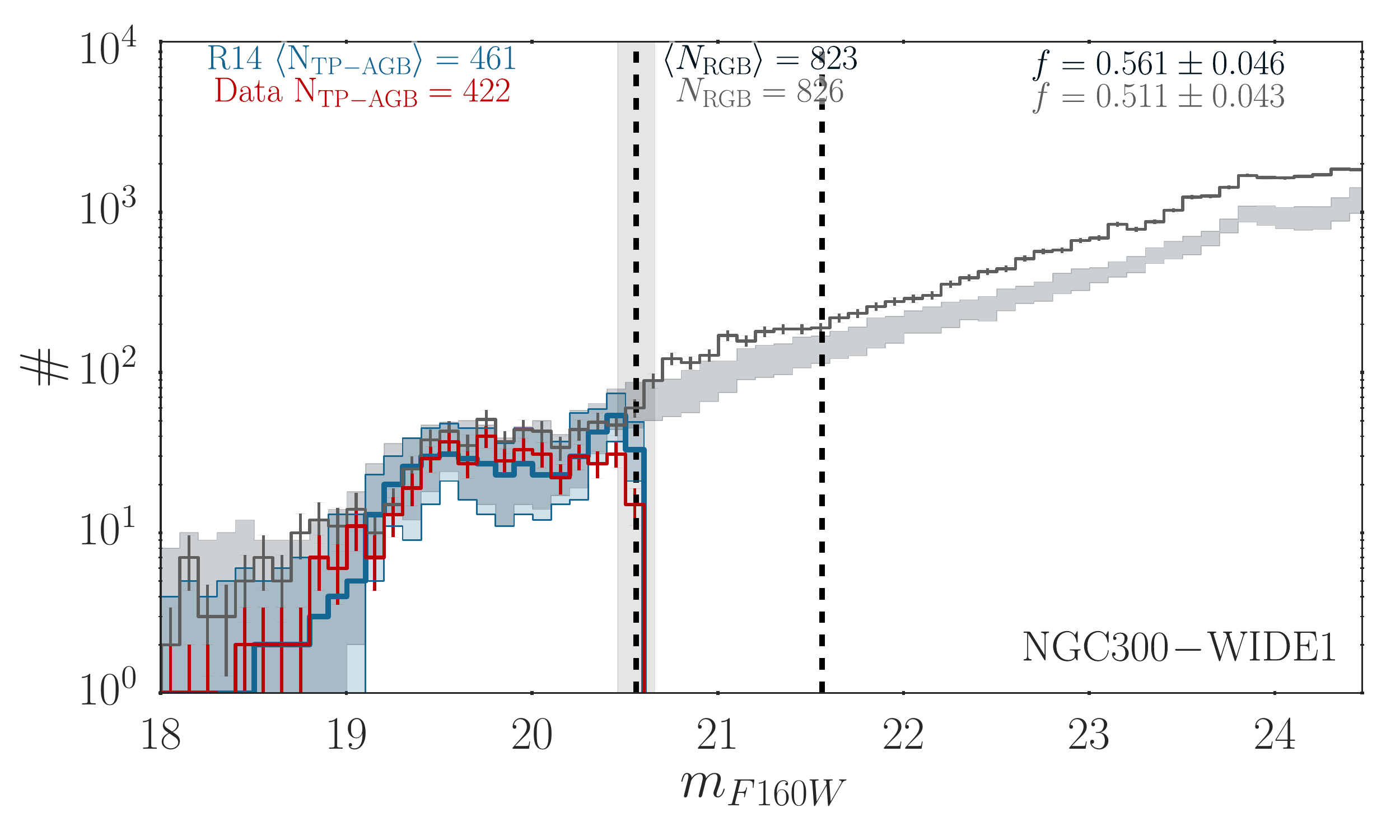}\\
\includegraphics[width=0.49\textwidth]{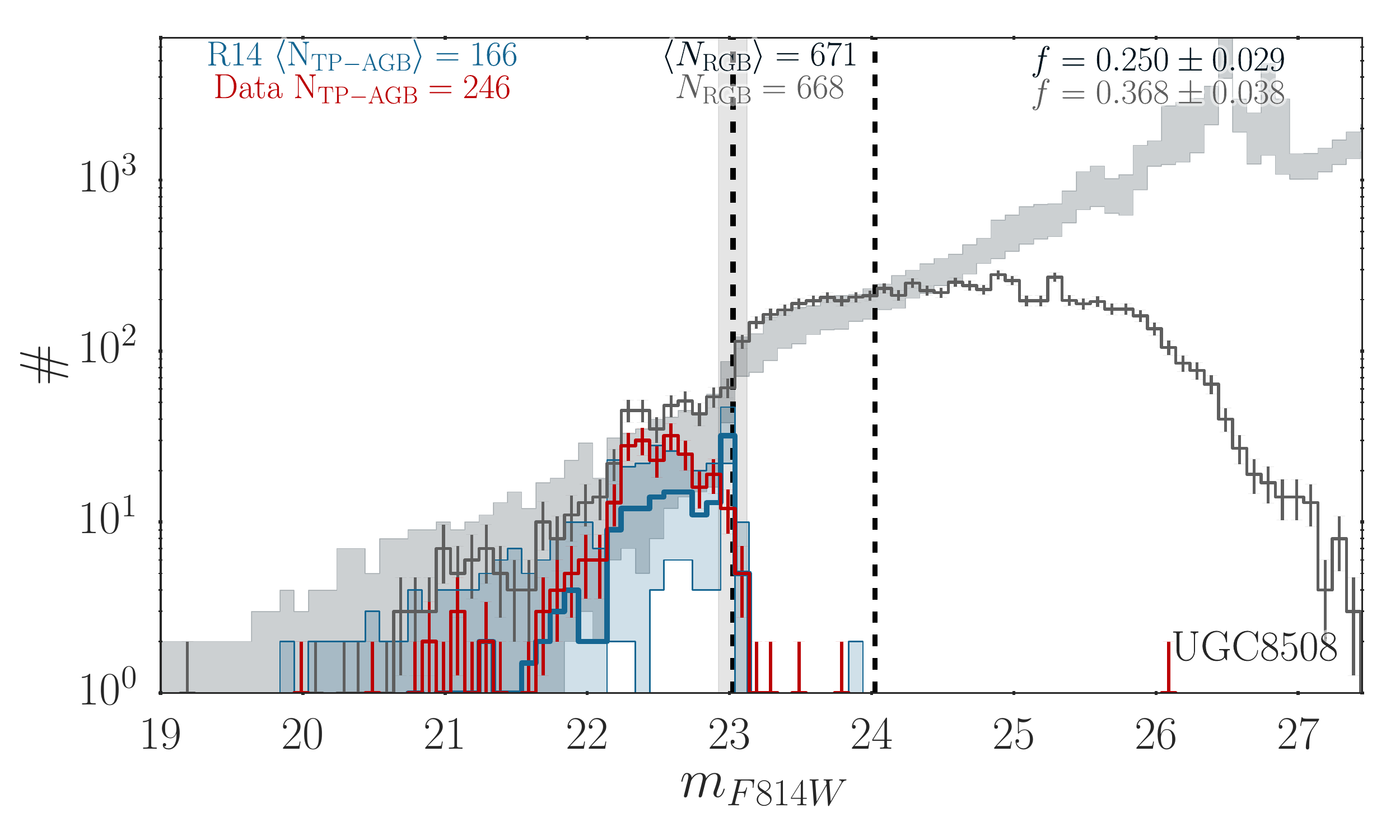}
\includegraphics[width=0.49\textwidth]{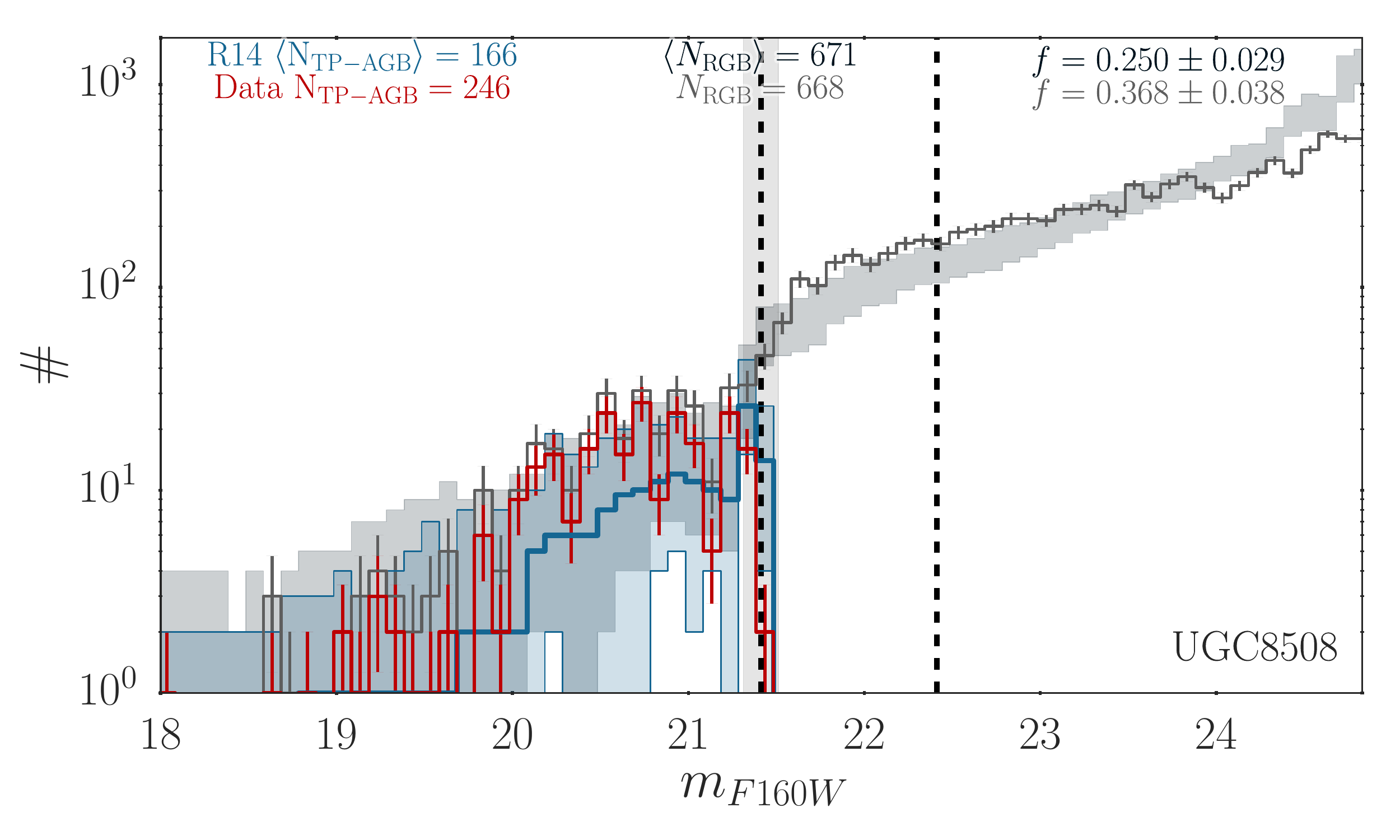}\\
\includegraphics[width=0.49\textwidth]{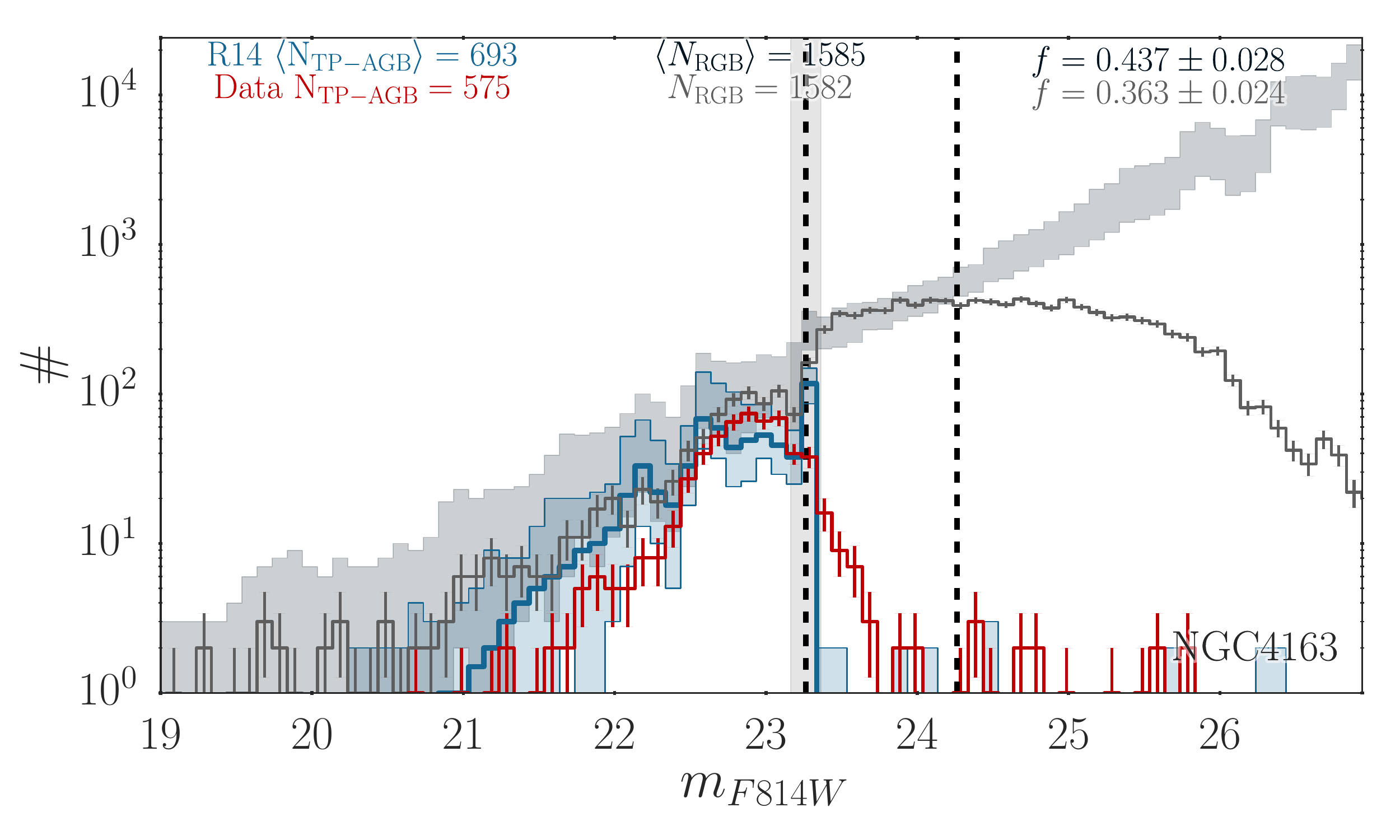}
\includegraphics[width=0.49\textwidth]{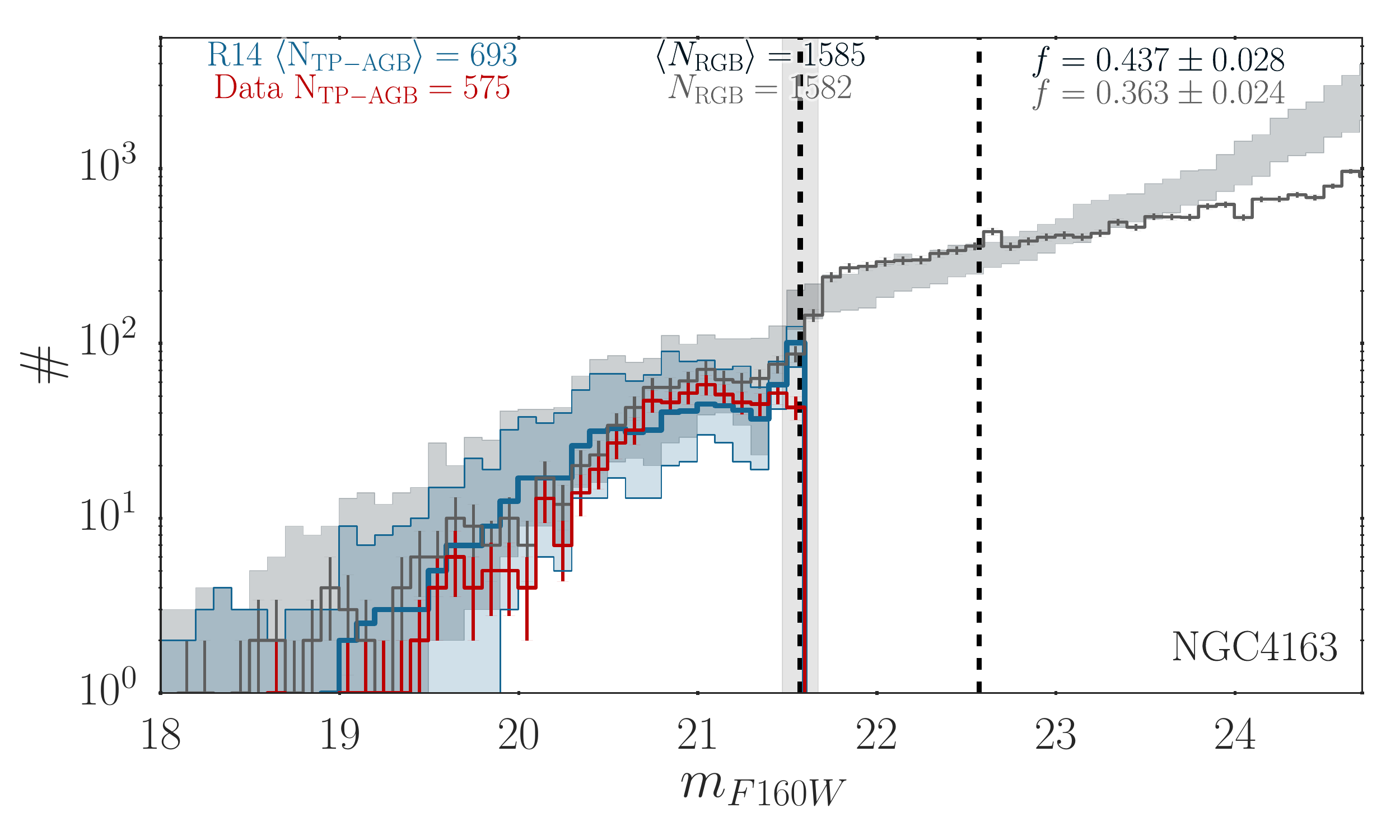}
\caption{Comparisons between the LF predicted by our models using the \citetalias{Rosenfield2014} TP-AGB tracks and the \citetalias{Melbourne2012} data. The data are marked by the dark gray line with Poisson uncertainties, with a separate red curve for the TP-AGB stars only. The models are derived from at least 100 {\tt TRILEGAL} simulations sampling the random uncertainties in the SFH: they are marked by gray shaded region denoting the 67\% confidence level interval. The contribution by TP-AGB stars is shown in blue with the same confidence interval shaded. The data have been completeness-corrected with AGB-SNAP ASTs (the faint fall off of the optical LFs are due to the constraint of both NIR and optical detections in the \citetalias{Melbourne2012} data). Over-plotted are the TRGB with a 0.2 mag dispersion as well as the 1 mag limit below the TRGB, the magnitude limits of the RGB selection box (vertical lines). The top of each panel show individual entries in Table \ref{tab_fracs} to compare the data and model.  Optical and near-infrared LFs are presented in the left and right panels, respectively.}
\label{fig:lfs}
\end{figure*}

Model and observed LFs are shown in Figure \ref{fig:lfs}. The \citetalias{Melbourne2012} catalog LFs (dark gray) are completeness corrected using the ANGST and AGB-SNAP  ASTs and are shown with Poisson uncertainties calculated for each magnitude bin. The $\sim100$ model LFs are combined and shown as 67\% confidence level intervals shaded in gray. The TP-AGB stars, identified by the method discussed in Section \ref{sec_ids}, are shown for the data (red) and for the model (blue) with the same confidence level intervals as the full LF (shaded blue). Vertical lines in each plot show the TRGB and 0.2 mag region around the TRGB excluded from the \narratio\ calculations (see Section \ref{sec_ids}). The fainter vertical line outlines the RGB box magnitude (but not color) limits used to scale the {\tt TRILEGAL} simulations. Finally, the top of each panel shows the listing from Table \ref{tab_fracs} of the number of TP-AGB and RGB stars observed and modeled as well as the corresponding \narratio\ ratios.

The LFs all follow the same trends, nearly a power-law in the F160W at magnitudes faintward of the TRGB, and some degree of agreement and spread in the TP-AGB region (brightward of the TRGB). The departure from a power-law of the observations at faint magnitudes, especially in the optical (despite being completeness corrected), is due to stars not detected in both F814W and F160W in the \citetalias{Melbourne2012} four-filter matched catalogs. In nearly every galaxy in both filters, the spread in the model TP-AGB LF from random SFH uncertainties is consistent with the data TP-AGB LF. The exceptions are the optical LFs of NGC~300-WIDE1, UGC~4305-1, and UGC~5139 where the model under-predicts the number of TP-AGB stars $\sim0.5$ mag brighter than the TRGB.

\subsection{Flux Contribution of TP-AGB Stars}
One issue quantified by in the \citetalias{Melbourne2012} study was the over-production of TP-AGB stars using the Padua models outside of the MCs. With the \citetalias{Rosenfield2014} models, it is worth revisiting the issue of the flux contribution of TP-AGB stars. Figure \ref{fig:m12} is a reproduction of Figure 7 of \citetalias{Melbourne2012} (gray) with our results over plotted (blue). Perfect agreement between data and model would be a ratio equaling 1, which is shown as a black solid horizontal line. Mean values for each model is shown as a dashed line. The \citet{Marigo2008} model (gray diamonds) emphasize the initial problem of extrapolating models calibrated in the MCs alone. On average, they over-predict both the number counts of TP-AGB stars (left panel) and the NIR ($F160W$) TP-AGB flux (right panel). The \citetalias{Girardi2010} models (gray squares), with the introduction of pre-dust mass loss, drastically reduce the offsets apparent in the \citet{Marigo2008} model. In contrast, the \citetalias{Rosenfield2014} models show the best agreement on average for both the observed number counts of TP-AGB stars (0.80) as well as the TP-AGB flux (0.90).

Finally, we present the relative age and mass distributions of TP-AGB stars in Figure \ref{fig:tpagb_dist} and a more detailed distribution of the simulated TP-AGB mass distributions in Figure \ref{fig:tpagb_masses}.

\begin{deluxetable}{lccccr}
\tablecaption{\narratio\ Comparisons\label{tab_fracs}}
\tablehead{
\colhead{Target} & \colhead{data} & \colhead{\citetalias{Rosenfield2014}} & \colhead{data} & \colhead{\citetalias{Rosenfield2014}} & \colhead{Fractional} \\
\colhead{} & \colhead{$\rm{N_{TP-AGB}}$} & \colhead{$\rm{N_{TP-AGB}}$} & \colhead{\narratiot} & \colhead{\narratiot} & \colhead{Difference}
}
\startdata
NGC~300-WIDE1 & $422\pm21$ & $462\pm21$ & $0.511\pm0.043$ & $0.560\pm0.046$ & $-0.097\pm0.016$ \\
UGC~8508 & $246\pm16$ & $167\pm13$ & $0.368\pm0.038$ & $0.249\pm0.029$ & $0.325\pm0.071$ \\
NGC~4163 & $575\pm24$ & $693\pm26$ & $0.363\pm0.024$ & $0.437\pm0.028$ & $-0.203\pm0.026$ \\
NGC~2403-DEEP & $181\pm13$ & $150\pm12$ & $0.614\pm0.081$ & $0.508\pm0.071$ & $0.171\pm0.047$ \\
NGC~2403-HALO-6 & $136\pm12$ & $93\pm10$ & $0.604\pm0.092$ & $0.414\pm0.071$ & $0.315\pm0.102$ \\
NGC~3741 & $241\pm16$ & $213\pm15$ & $0.429\pm0.046$ & $0.377\pm0.042$ & $0.120\pm0.026$ \\
UGC~4305-1 & $549\pm23$ & $340\pm18$ & $0.558\pm0.042$ & $0.346\pm0.030$ & $0.380\pm0.061$ \\
UGC~4305-2 & $612\pm25$ & $602\pm25$ & $0.463\pm0.031$ & $0.454\pm0.031$ & $0.019\pm0.003$ \\
UGC~4459 & $247\pm16$ & $327\pm18$ & $0.351\pm0.036$ & $0.465\pm0.043$ & $-0.324\pm0.063$ \\
UGC~5139 & $475\pm22$ & $210\pm15$ & $0.401\pm0.030$ & $0.178\pm0.017$ & $0.557\pm0.096$ \\
KDG~73 & $49\pm7$ & $30\pm5$ & $0.282\pm0.062$ & $0.171\pm0.044$ & $0.392\pm0.187$ \\
UGC~A292 & $71\pm8$ & $115\pm11$ & $0.477\pm0.096$ & $0.779\pm0.137$ & $-0.635\pm0.239$ \\
{\bf Mean} & $317\pm62$ & $283\pm58$ & $0.452\pm0.197$ & $0.412\pm0.200$ & $0.089\pm0.082$
\enddata
\tablecomments{The data and \citetalias{Rosenfield2014} \narratio. Both are calculated from the NIR matched catalogs (see section \ref{sec_ids}). The uncertainties reported on the mean are the quadrature of the listed uncertainties of each field and the mean fractional difference is calculated from the mean ratio values, not for example, the mean of the individual fractional differences.}
\end{deluxetable}

\begin{figure*}
\includegraphics[width=0.49\textwidth]{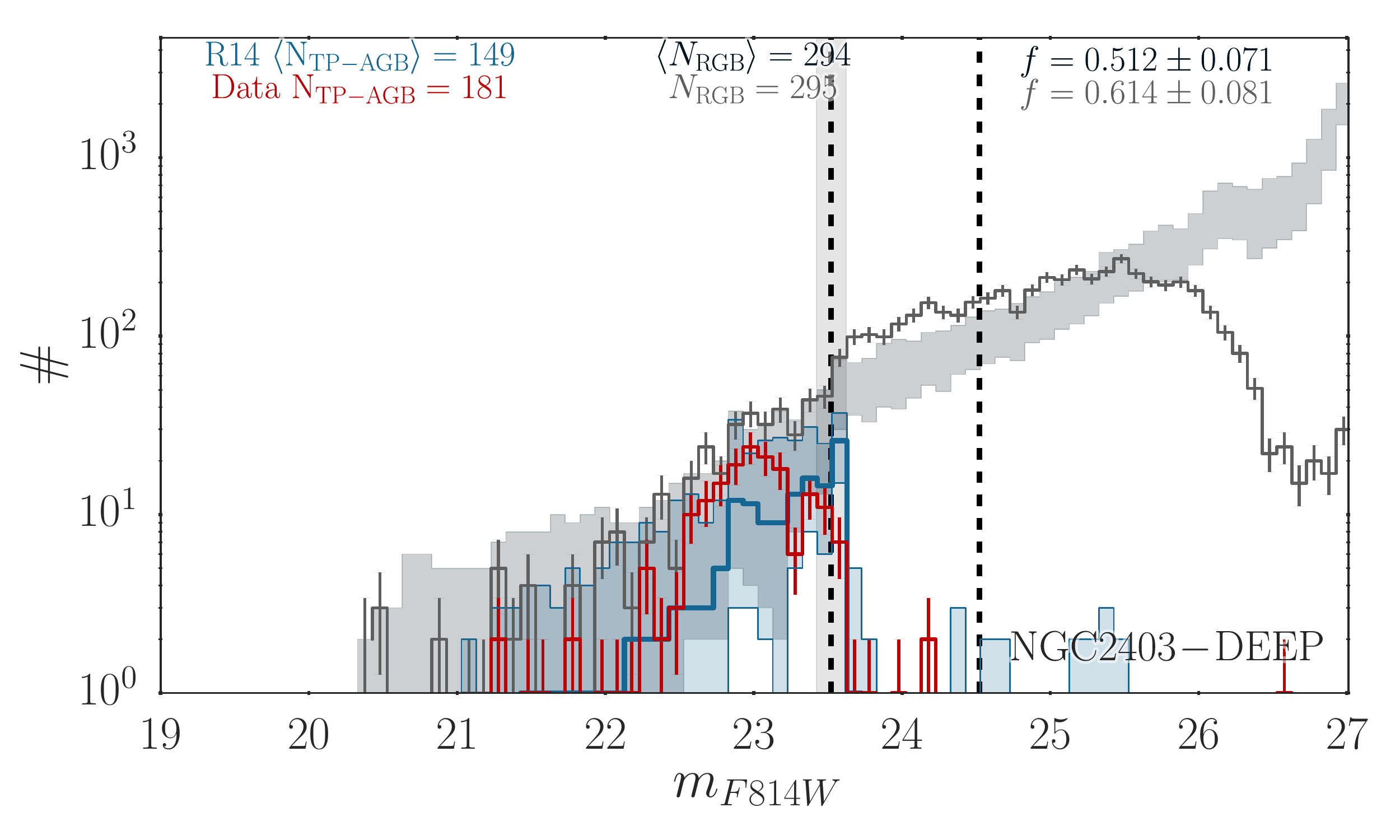}
\includegraphics[width=0.49\textwidth]{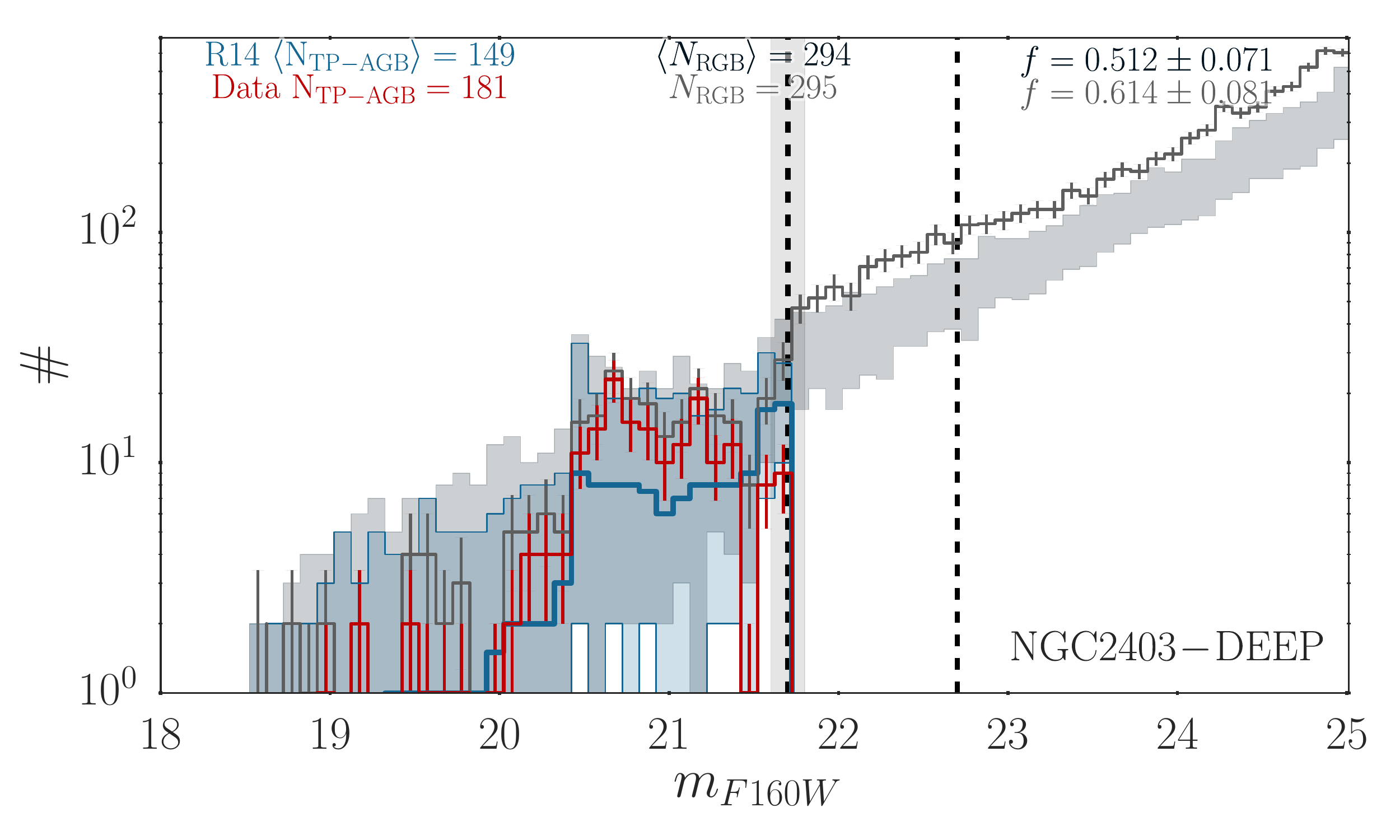}\\
\includegraphics[width=0.49\textwidth]{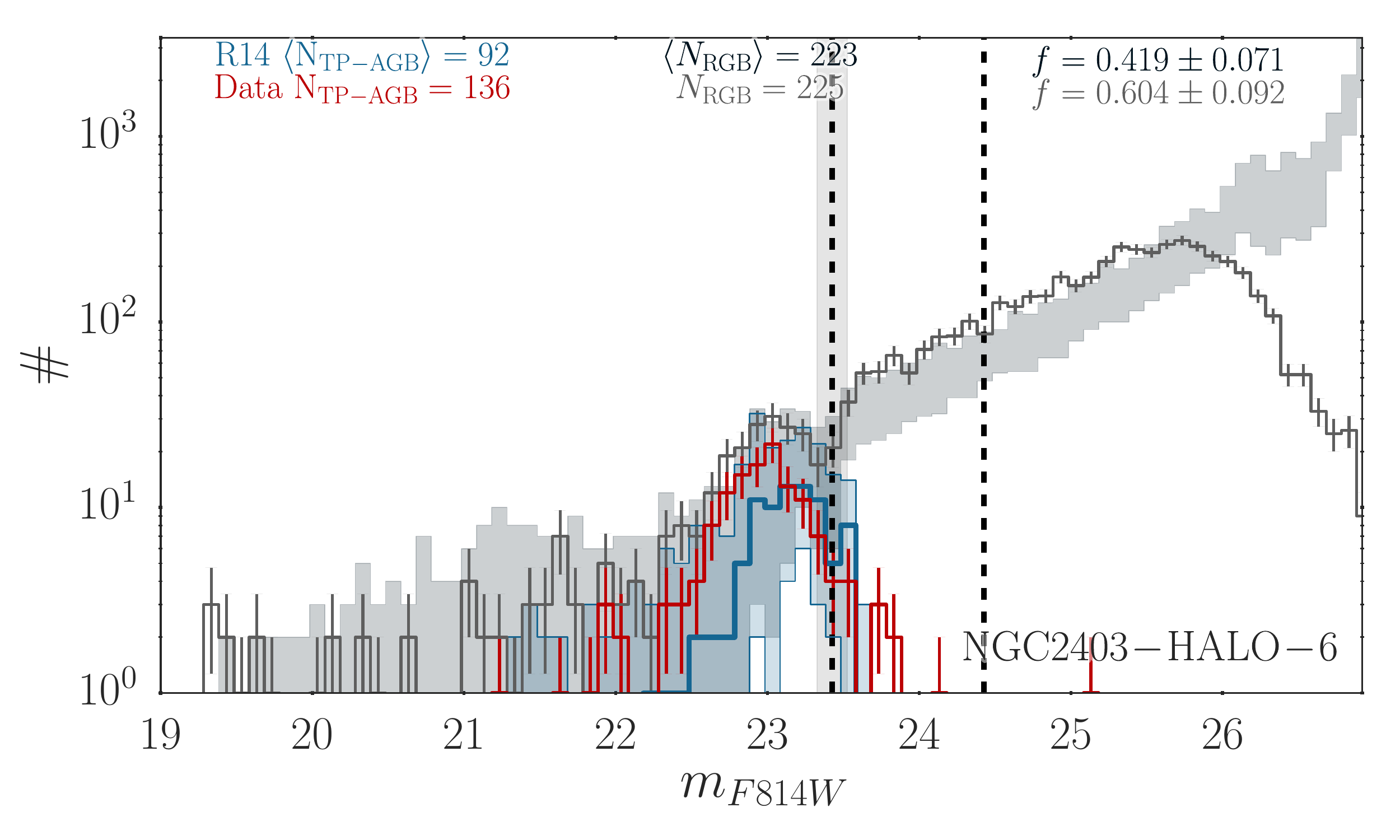}
\includegraphics[width=0.49\textwidth]{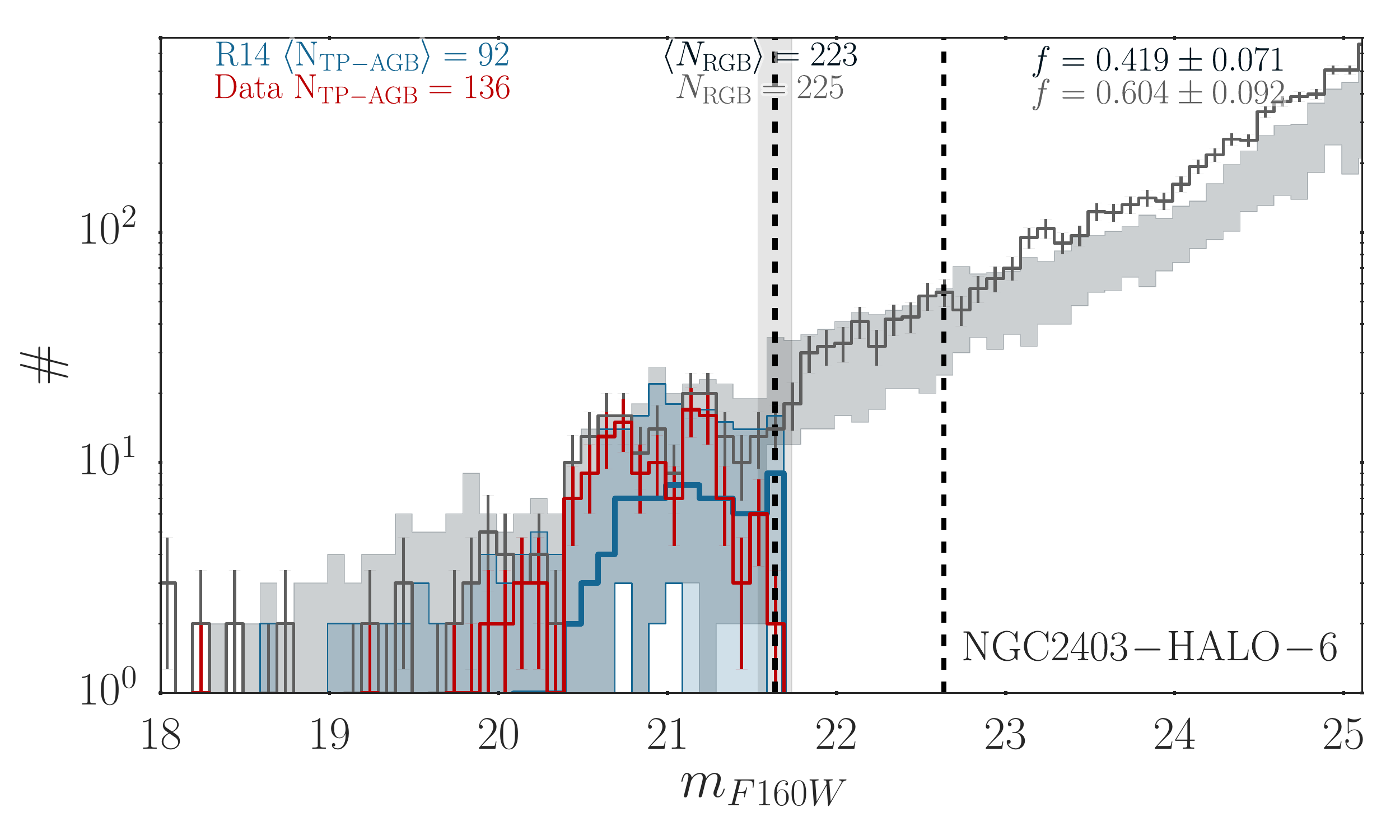}\\
\includegraphics[width=0.49\textwidth]{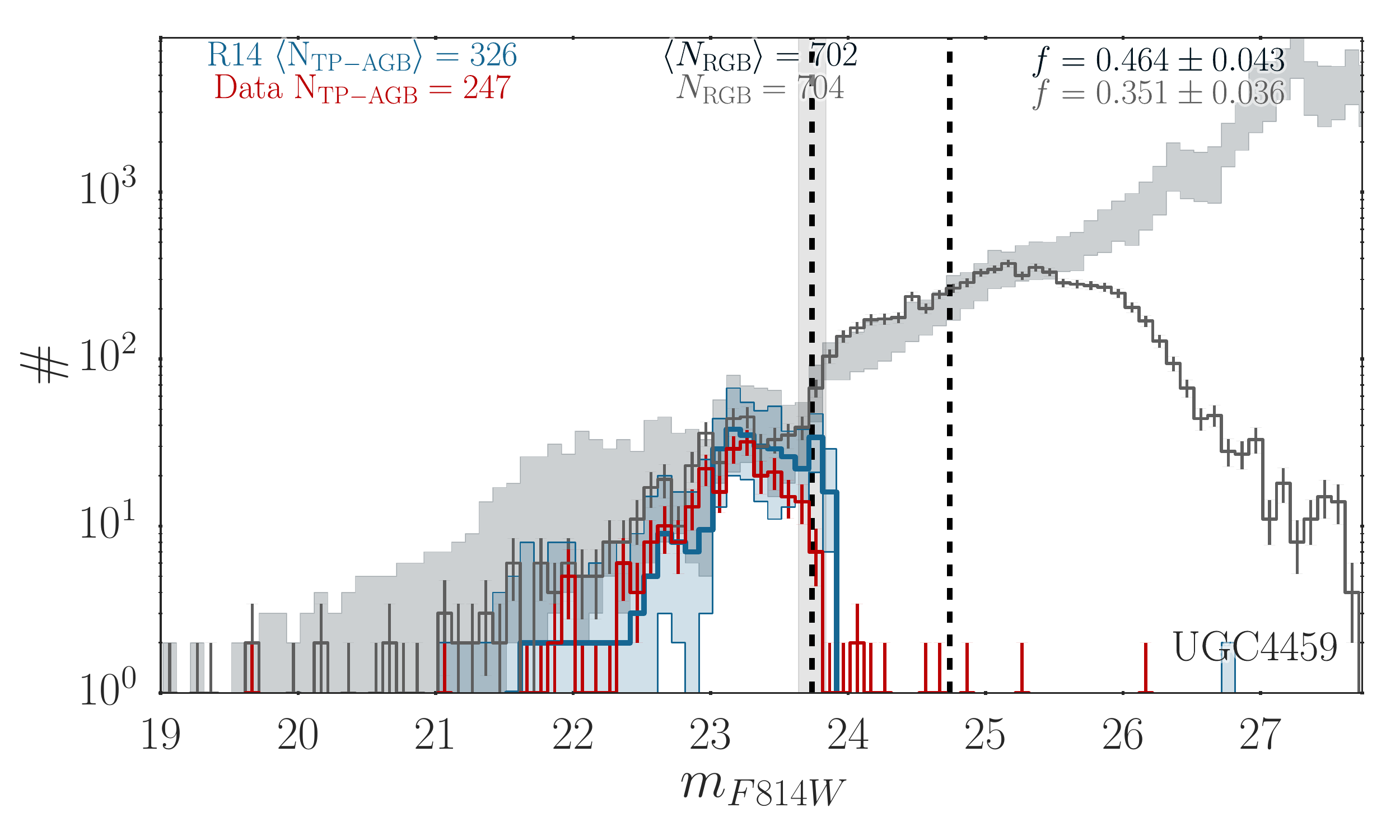}
\includegraphics[width=0.49\textwidth]{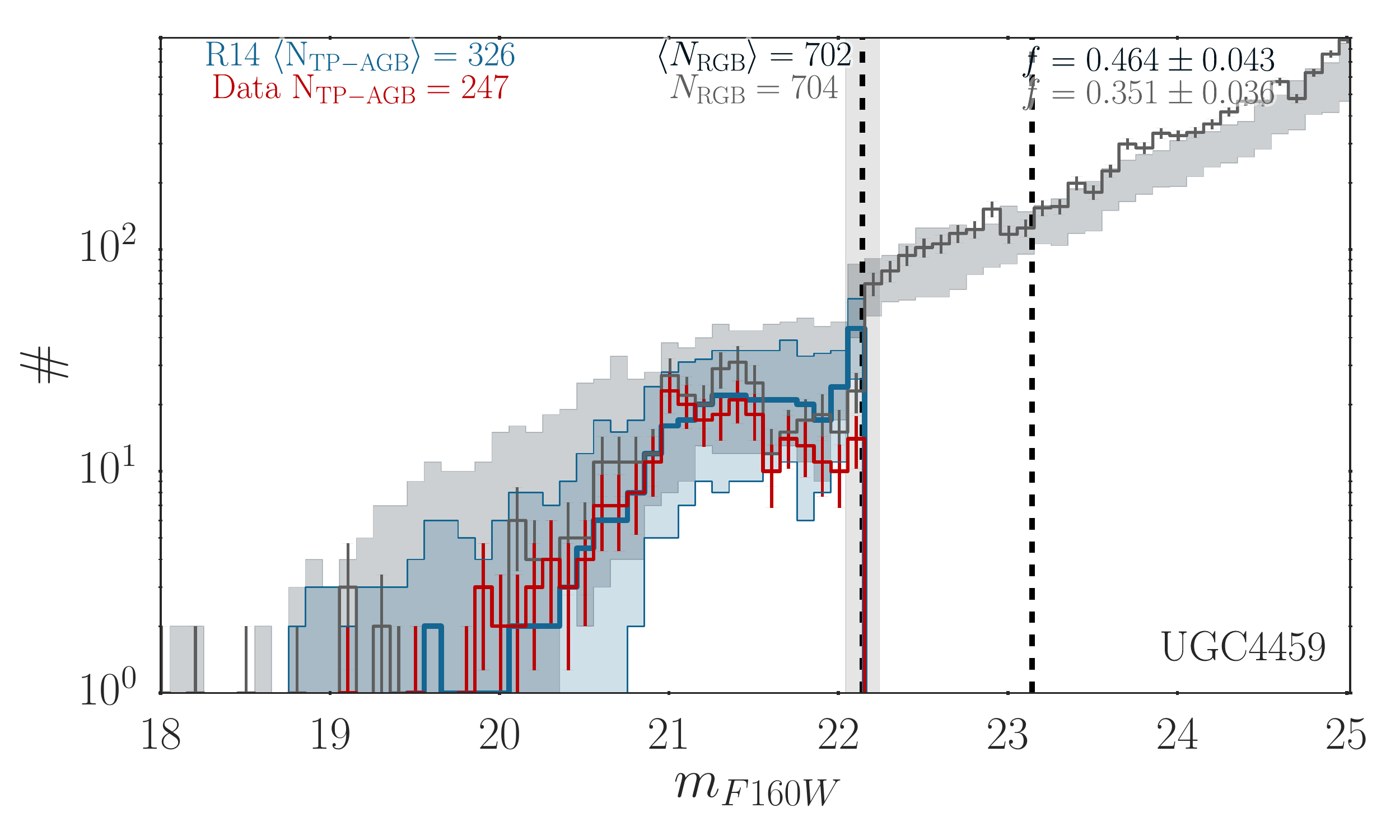}\\
\figurenum{\ref{fig:lfs} continued}
\caption{}
\end{figure*}

\begin{figure*}
\includegraphics[width=0.49\textwidth]{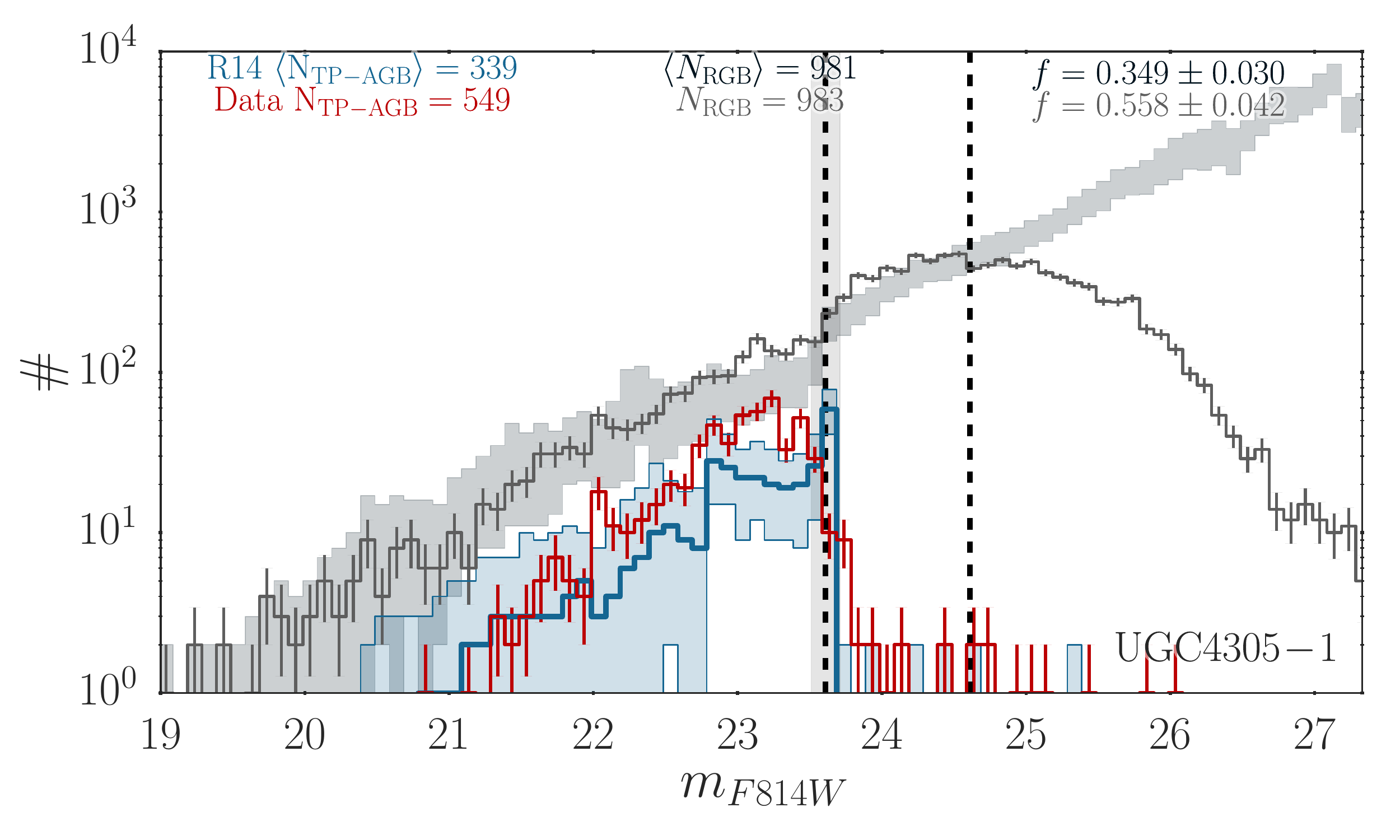}
\includegraphics[width=0.49\textwidth]{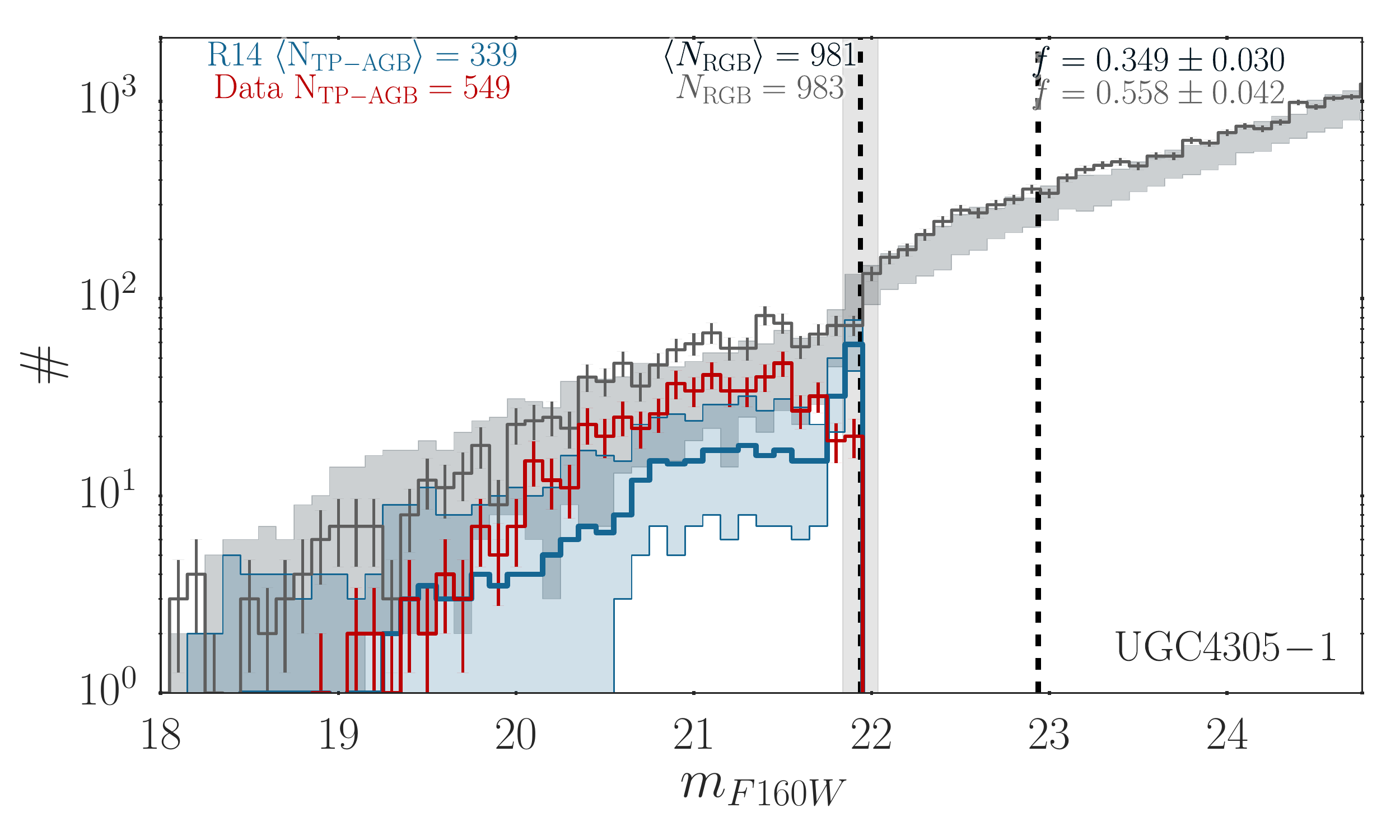}
\includegraphics[width=0.49\textwidth]{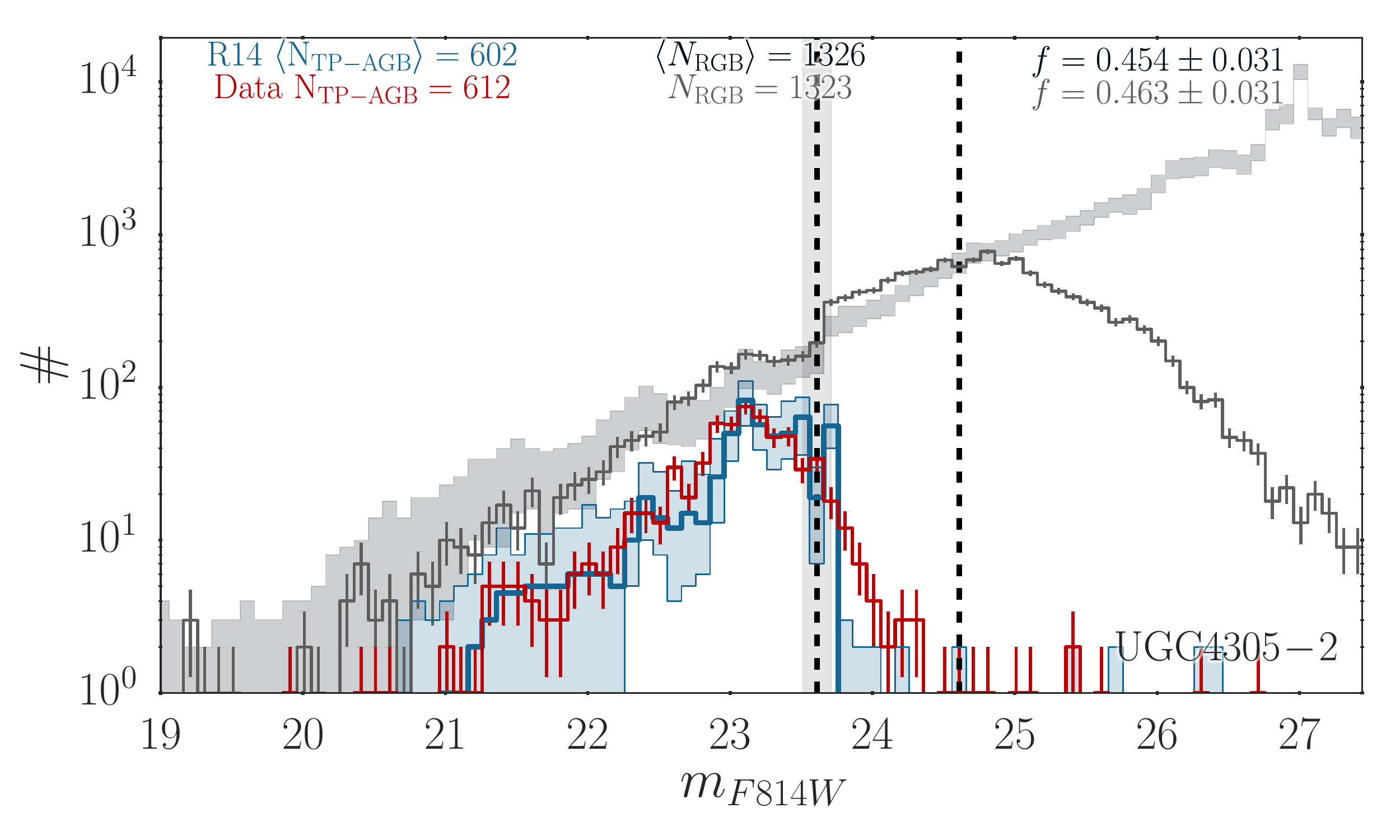}
\includegraphics[width=0.49\textwidth]{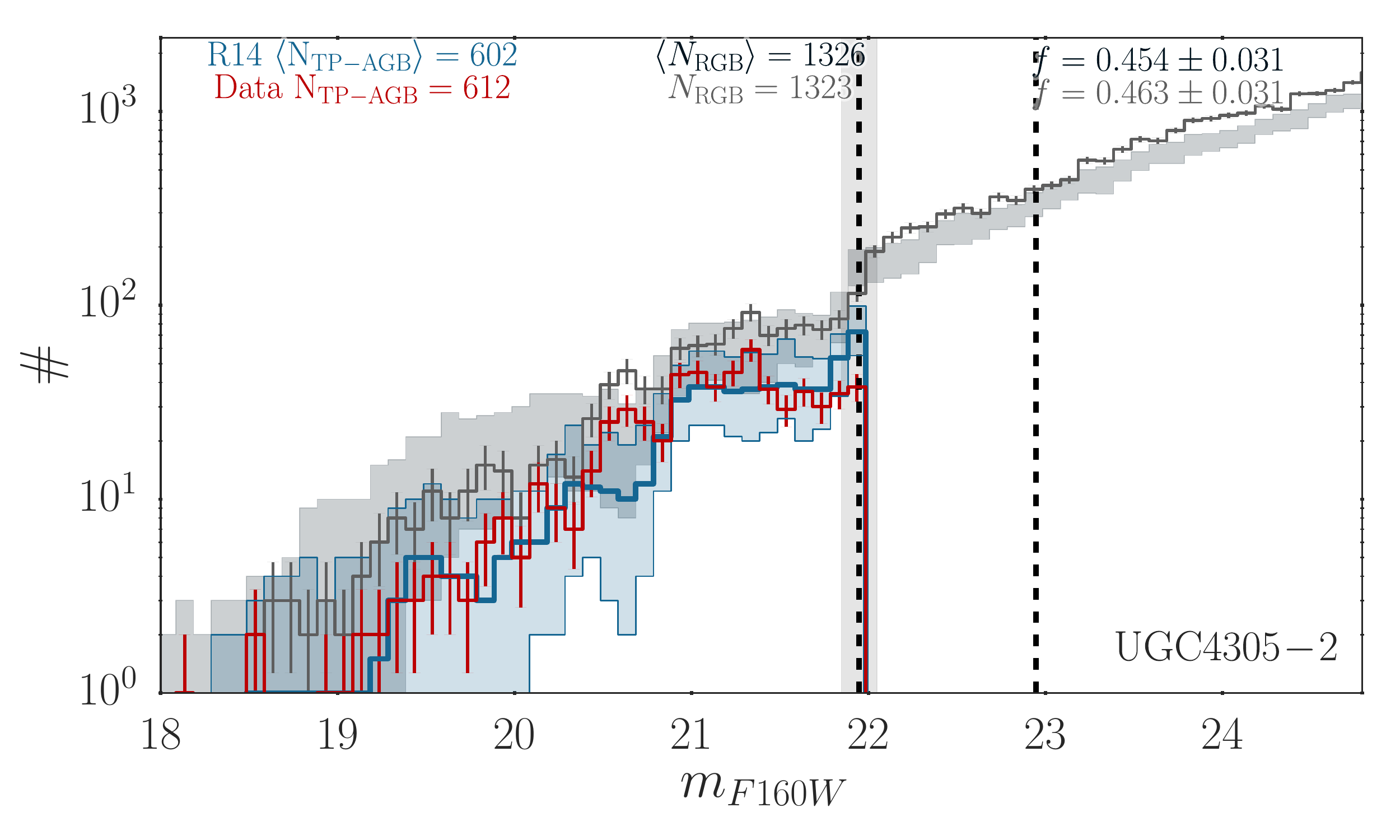}\\
\includegraphics[width=0.49\textwidth]{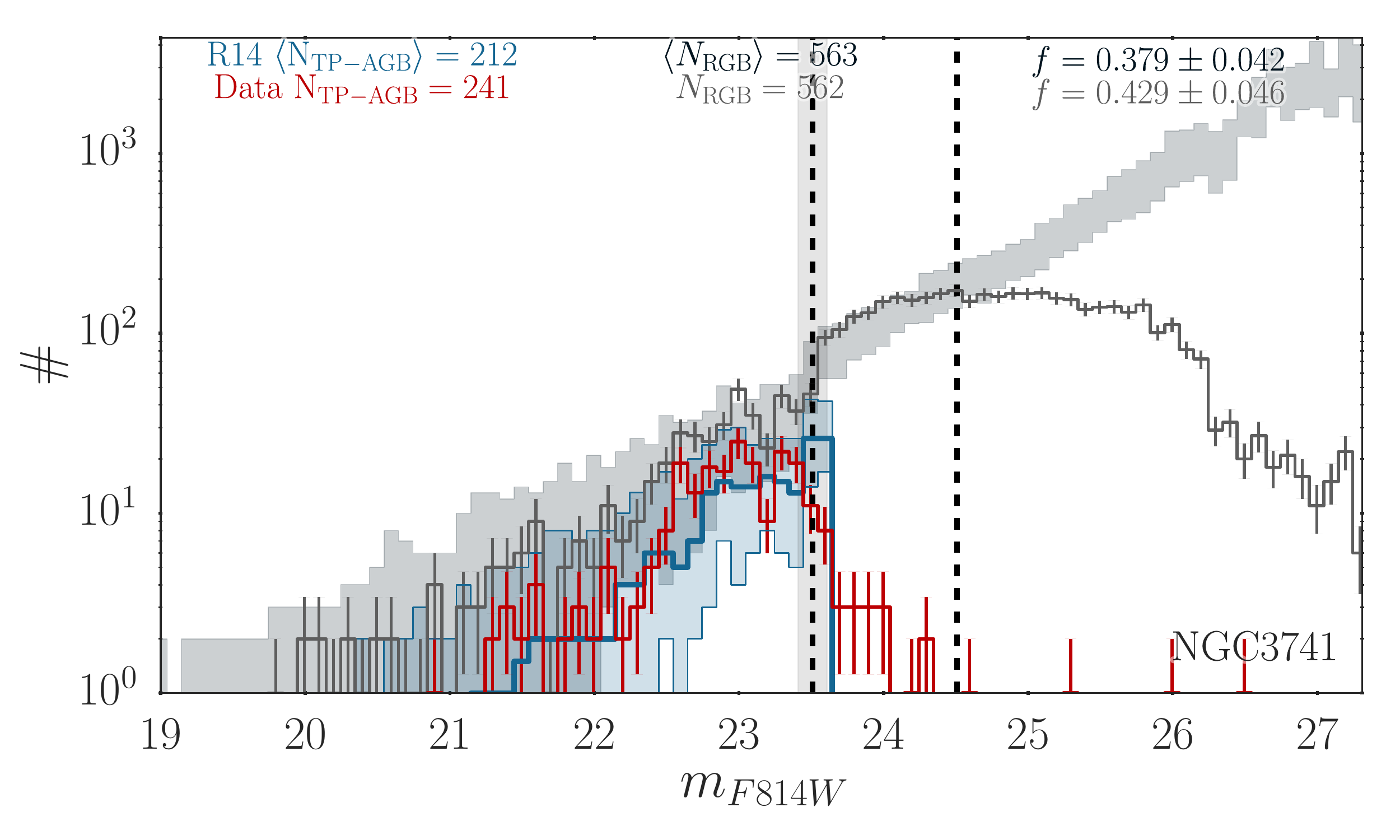}
\includegraphics[width=0.49\textwidth]{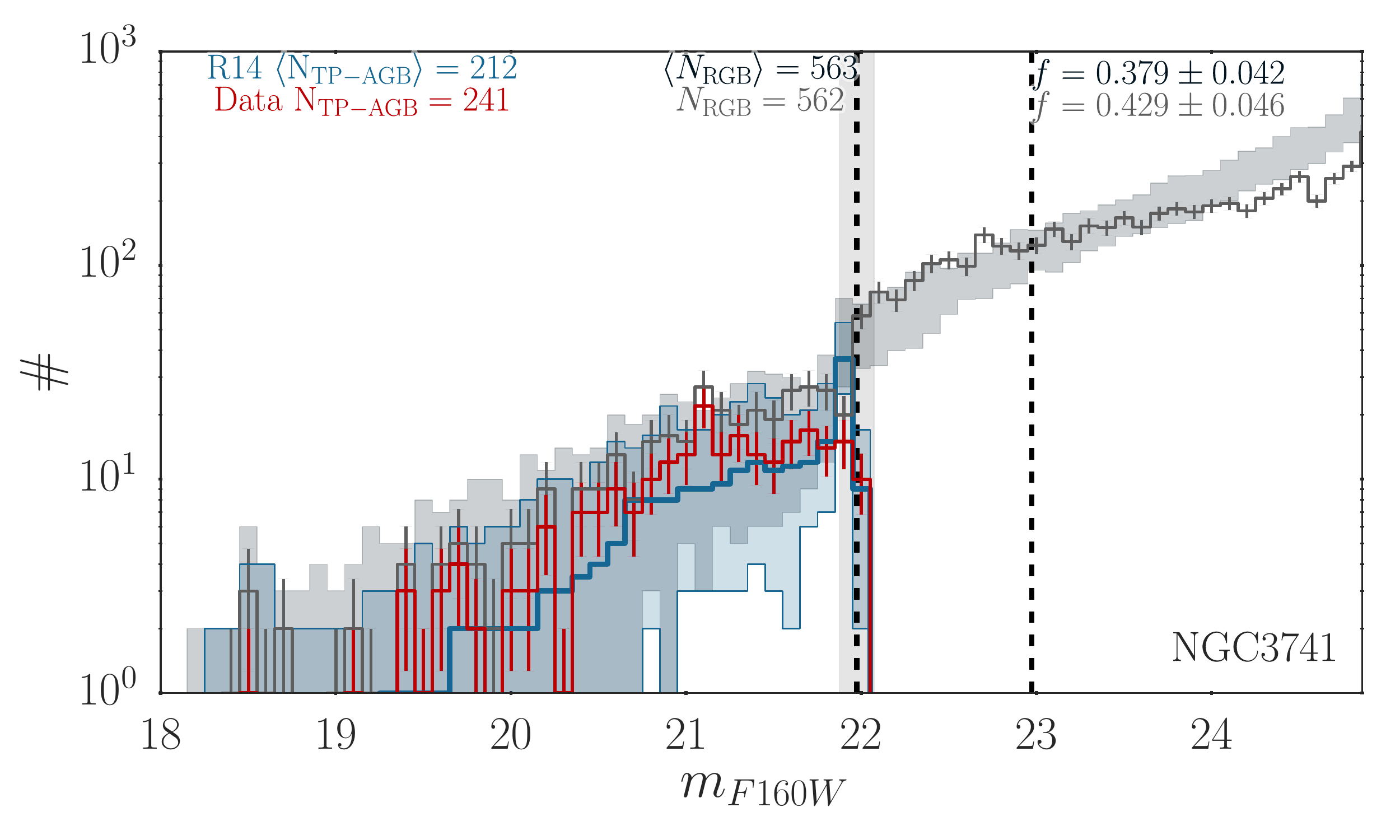}\\
\figurenum{\ref{fig:lfs} continued}
\caption{}
\end{figure*}

\begin{figure*}
\includegraphics[width=0.49\textwidth]{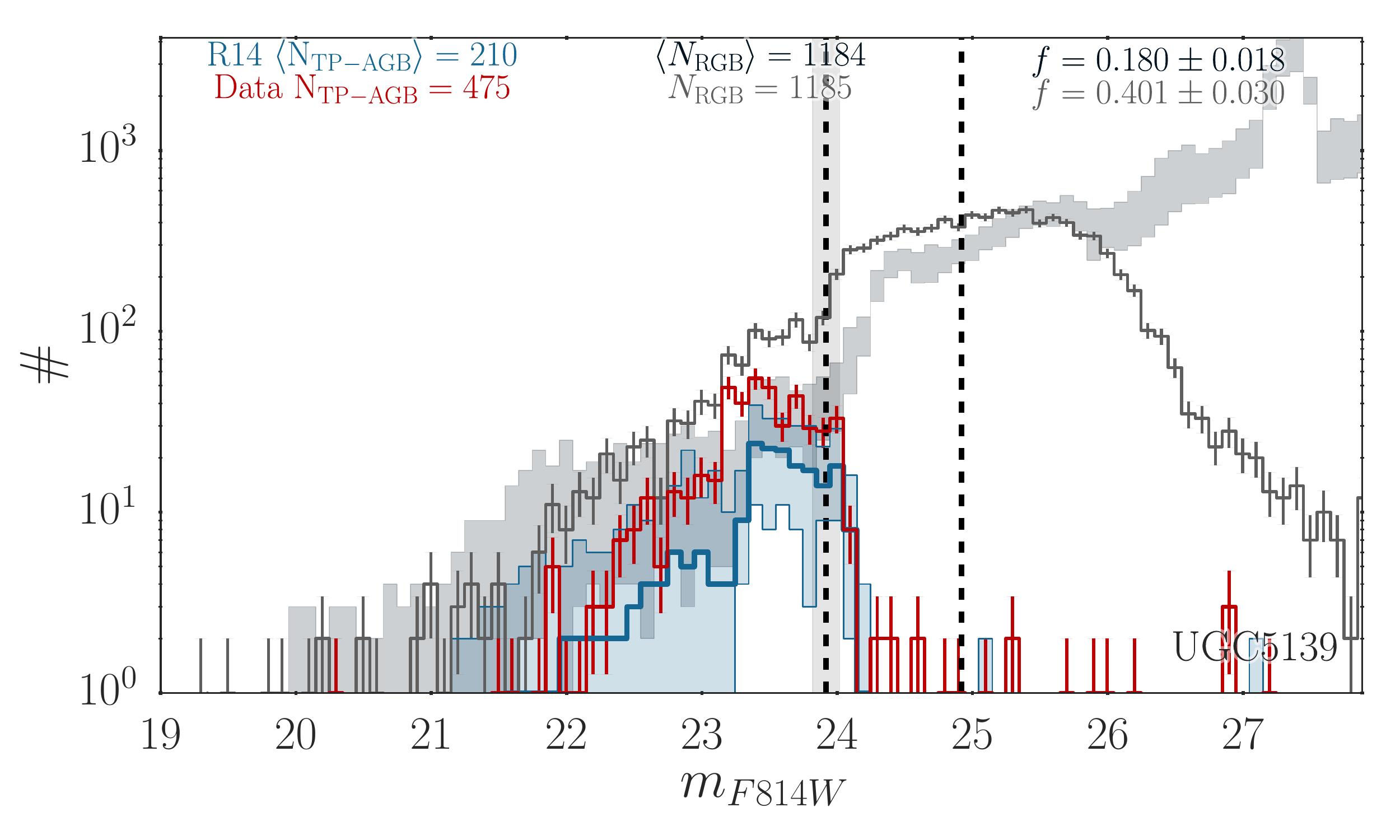}
\includegraphics[width=0.49\textwidth]{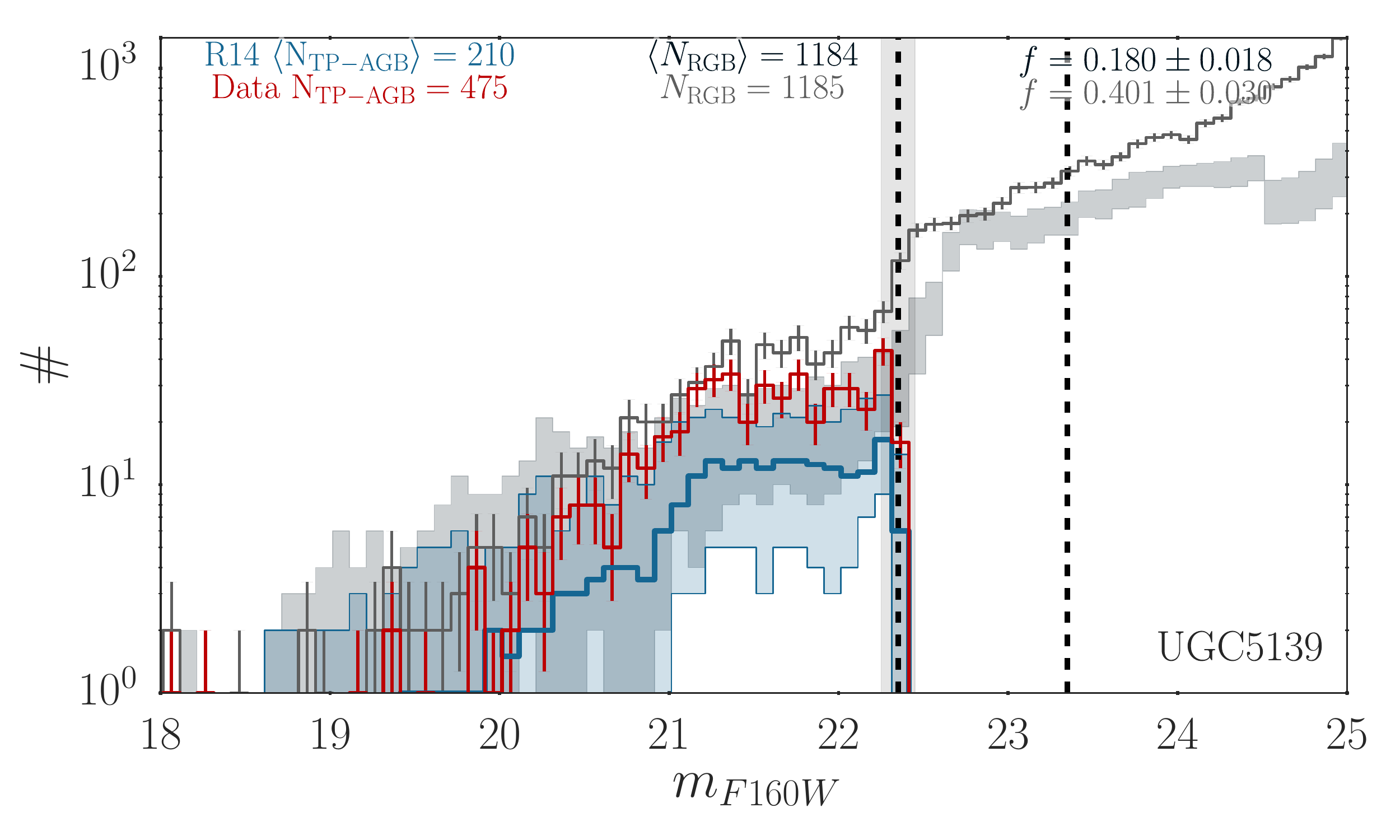}\\
\includegraphics[width=0.49\textwidth]{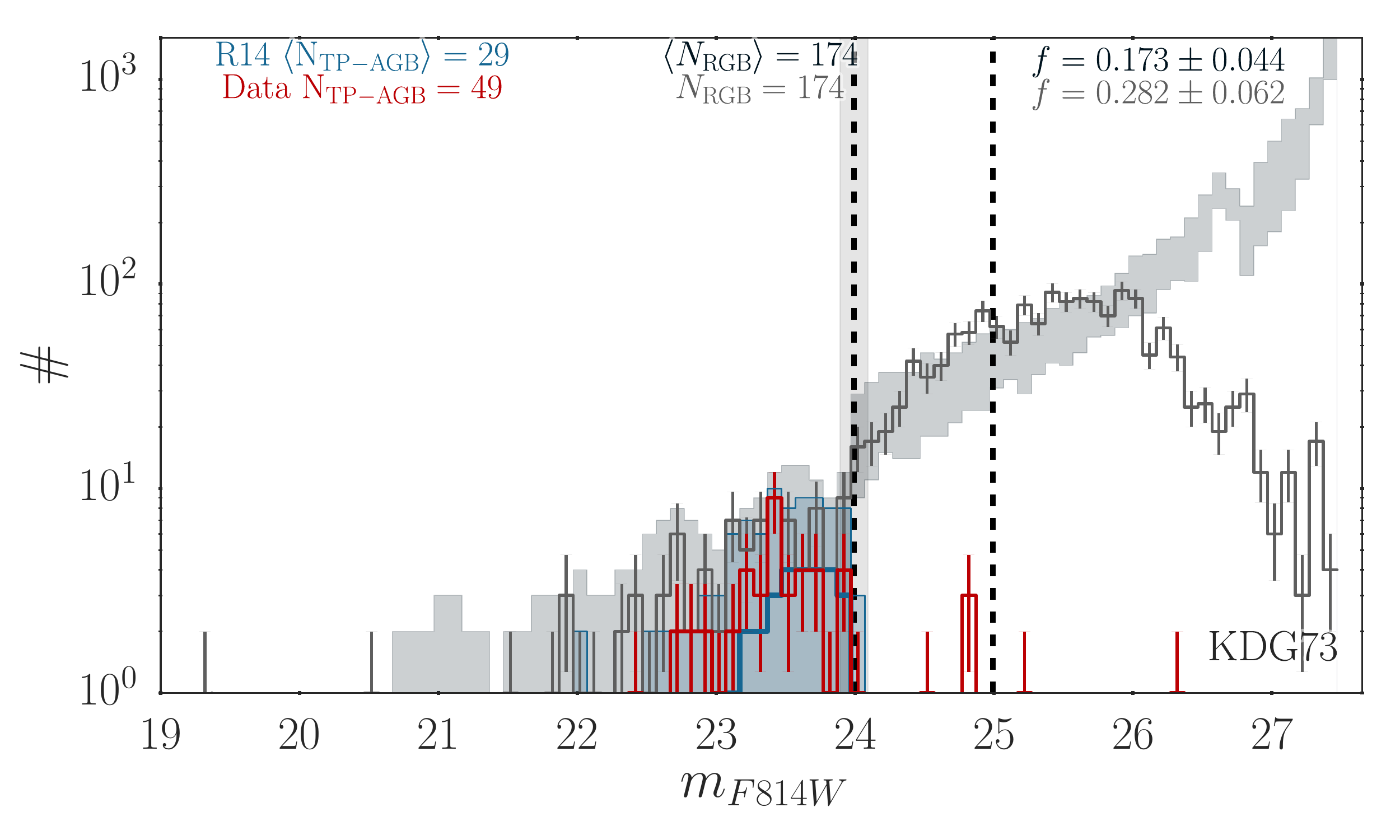}
\includegraphics[width=0.49\textwidth]{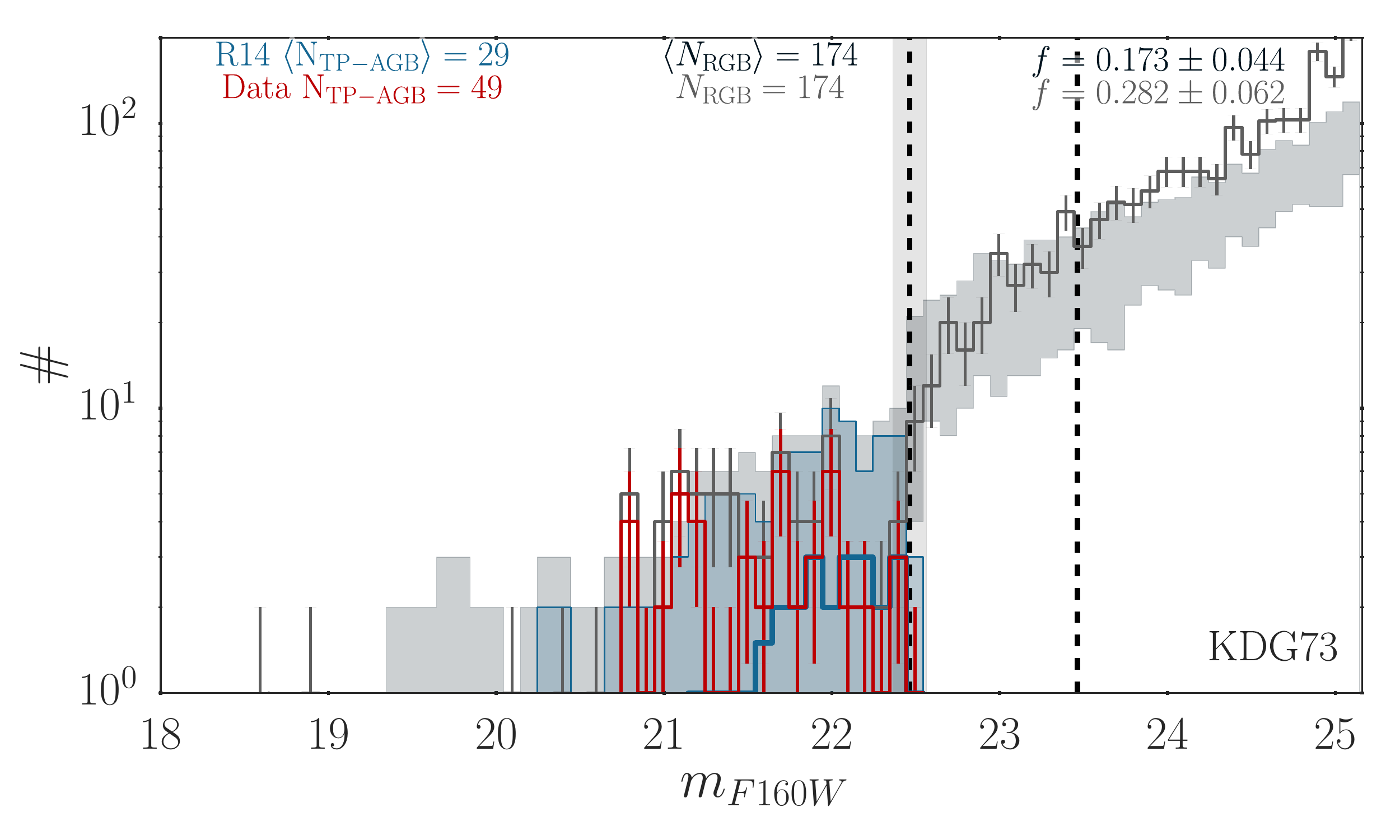}\\
\includegraphics[width=0.49\textwidth]{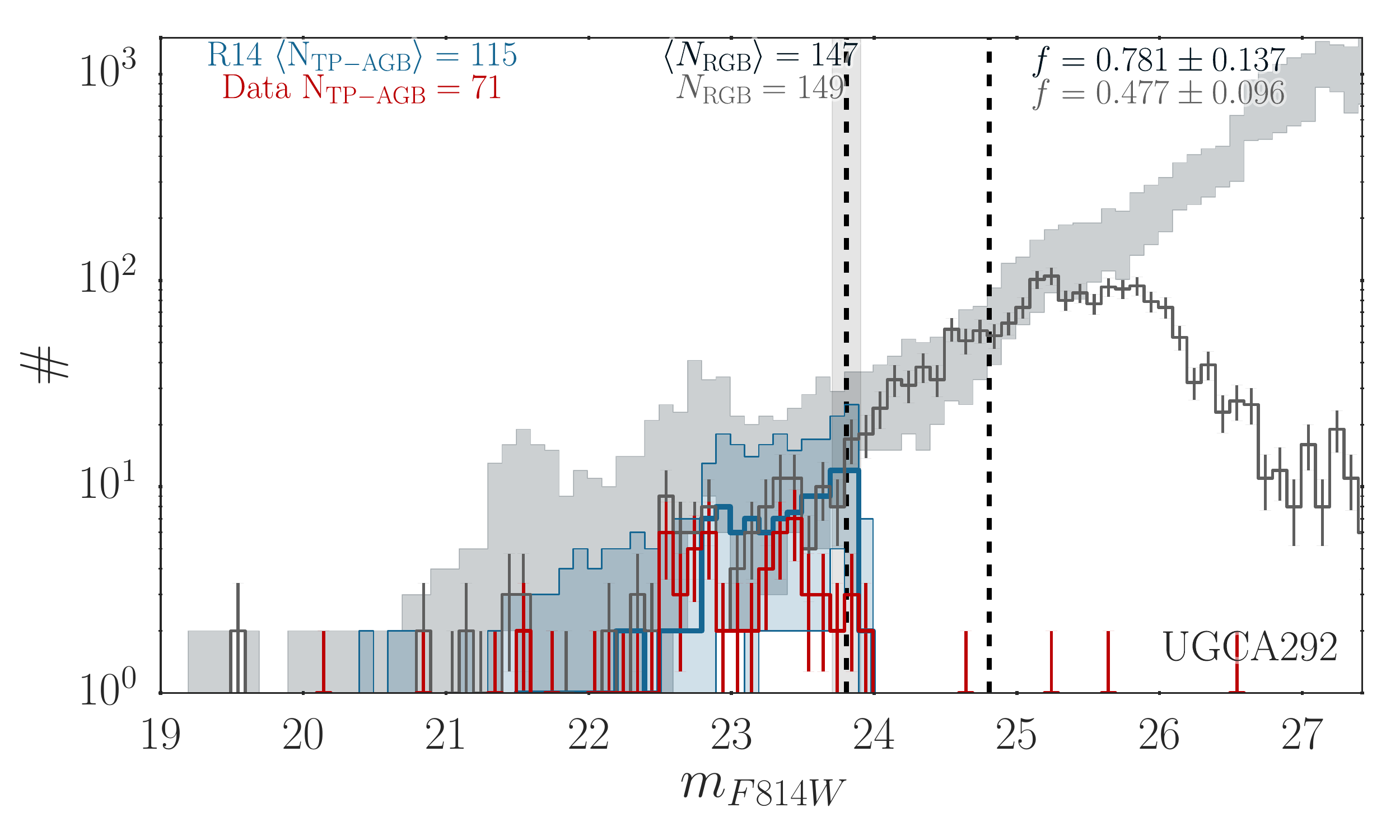}
\includegraphics[width=0.49\textwidth]{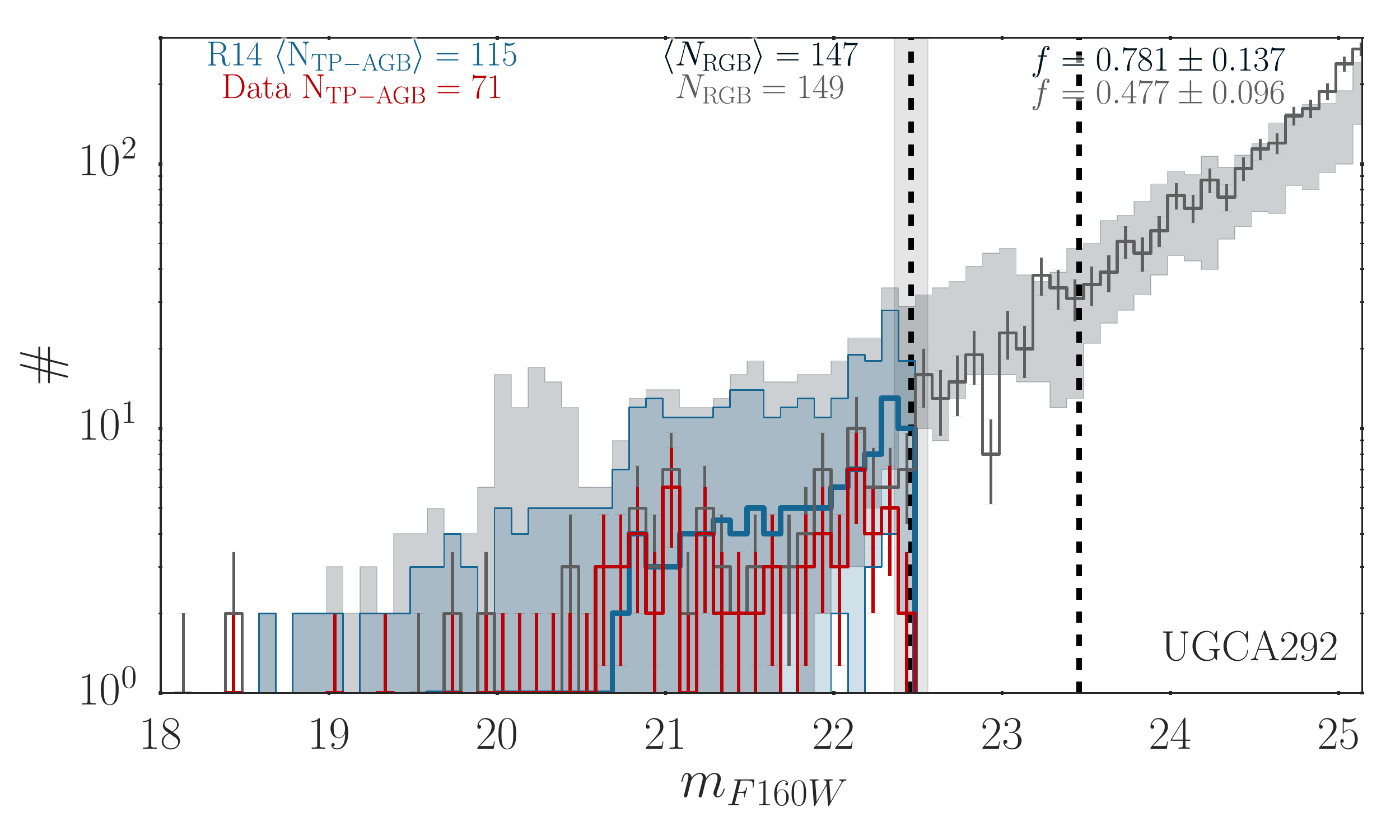}
\figurenum{\ref{fig:lfs} continued}
\caption{}
\end{figure*}

\begin{figure*}
\includegraphics[width=0.49\textwidth]{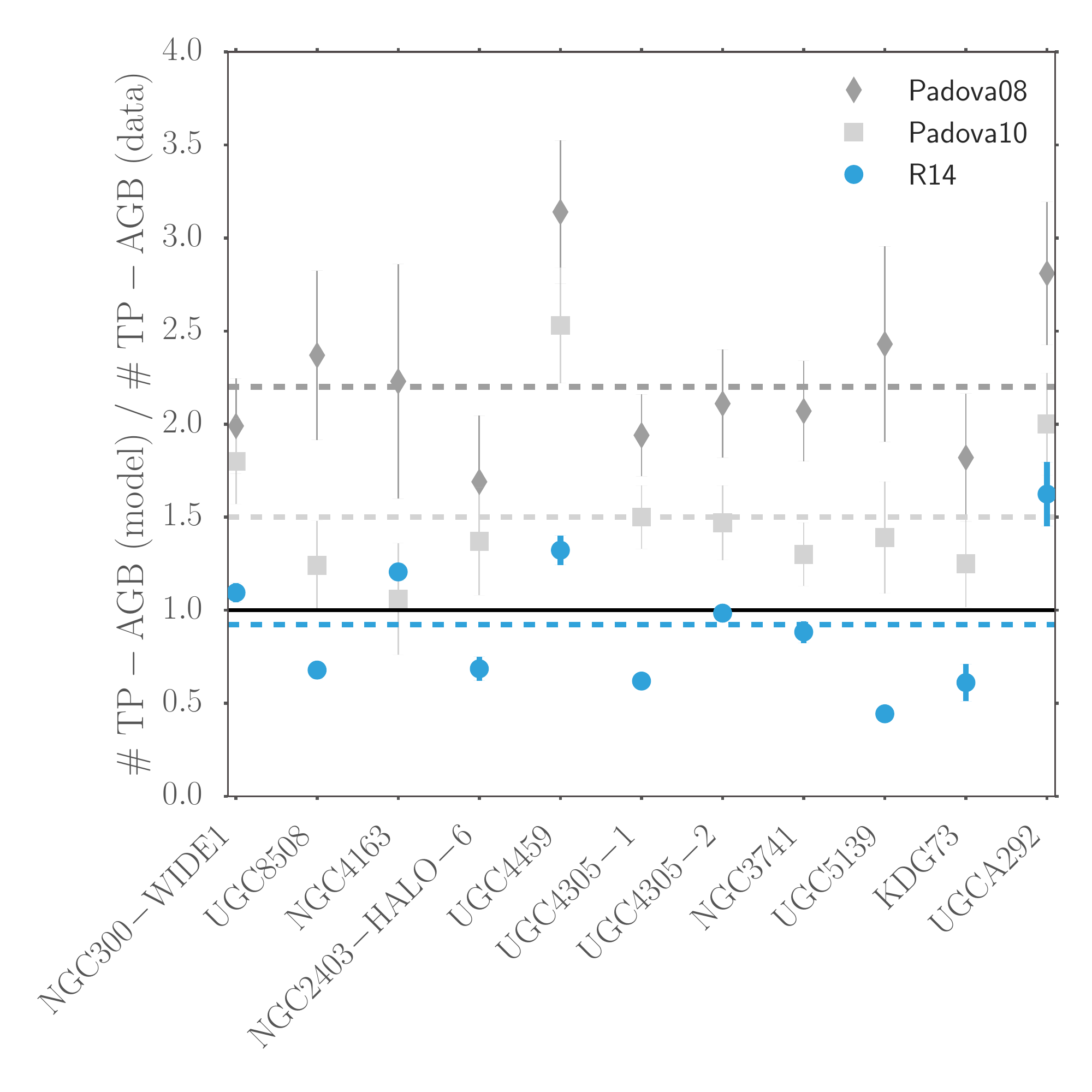}
\includegraphics[width=0.49\textwidth]{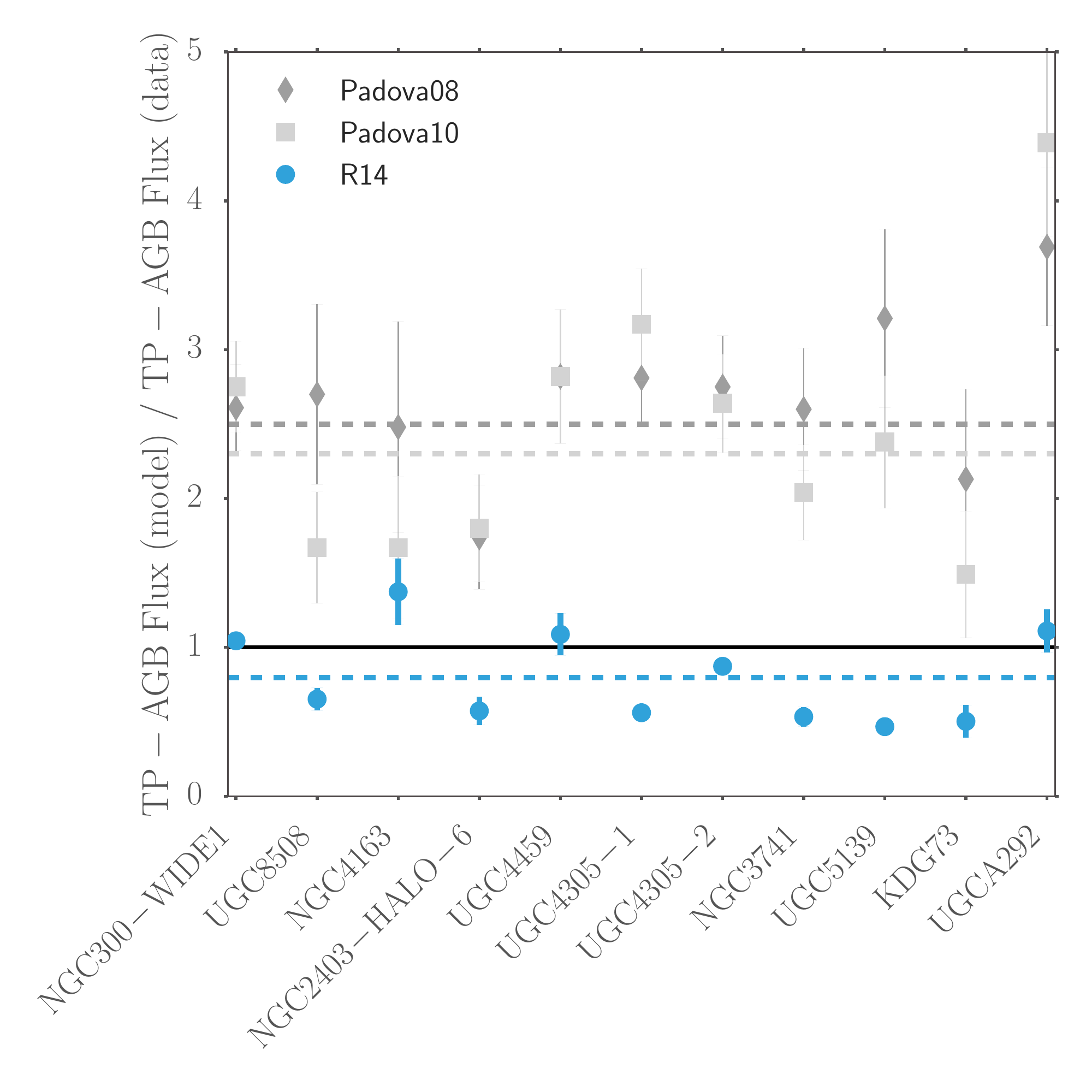}
\caption{\citetalias{Rosenfield2014} TP-AGB models improve the prediction the number counts of TP-AGB stars ({\it left panel}) as well as $F160W$ TP-AGB flux ({\it right panel}). Shown in comparison is a modified version of Figure 7 of \citetalias{Melbourne2012}: Padova 2008 \citep[gray diamonds;][]{Marigo2008} and Padova 2010 (gray squares; \citetalias{Girardi2010}), data and model are recalculated from \citetalias{Melbourne2012} tables with their weighted mean values shown as dashed gray horizontal lines. On average, galaxies in present study with \citetalias{Rosenfield2014} TP-AGB models (blue) show much better agreement to the data.}
\label{fig:m12}
\end{figure*}

\begin{figure*}
\includegraphics[width=\textwidth]{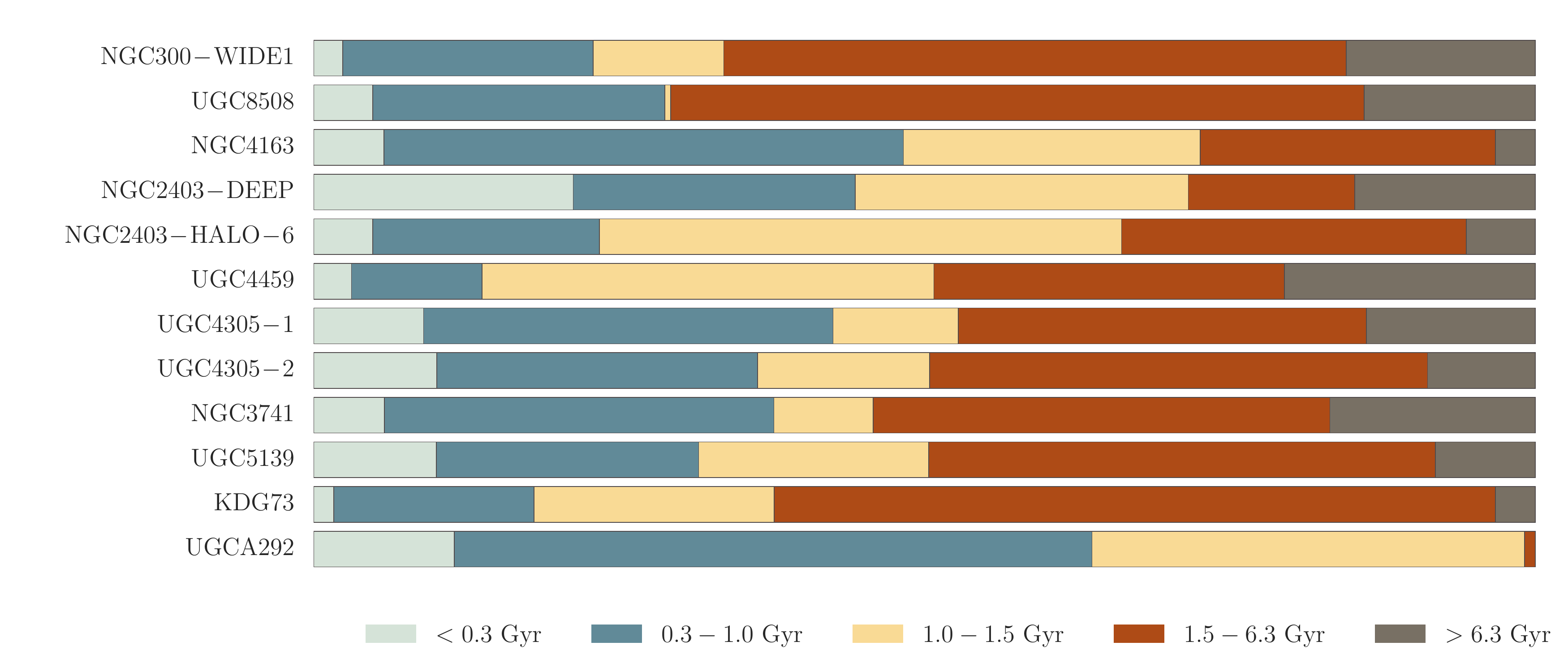} \\
\includegraphics[width=\textwidth]{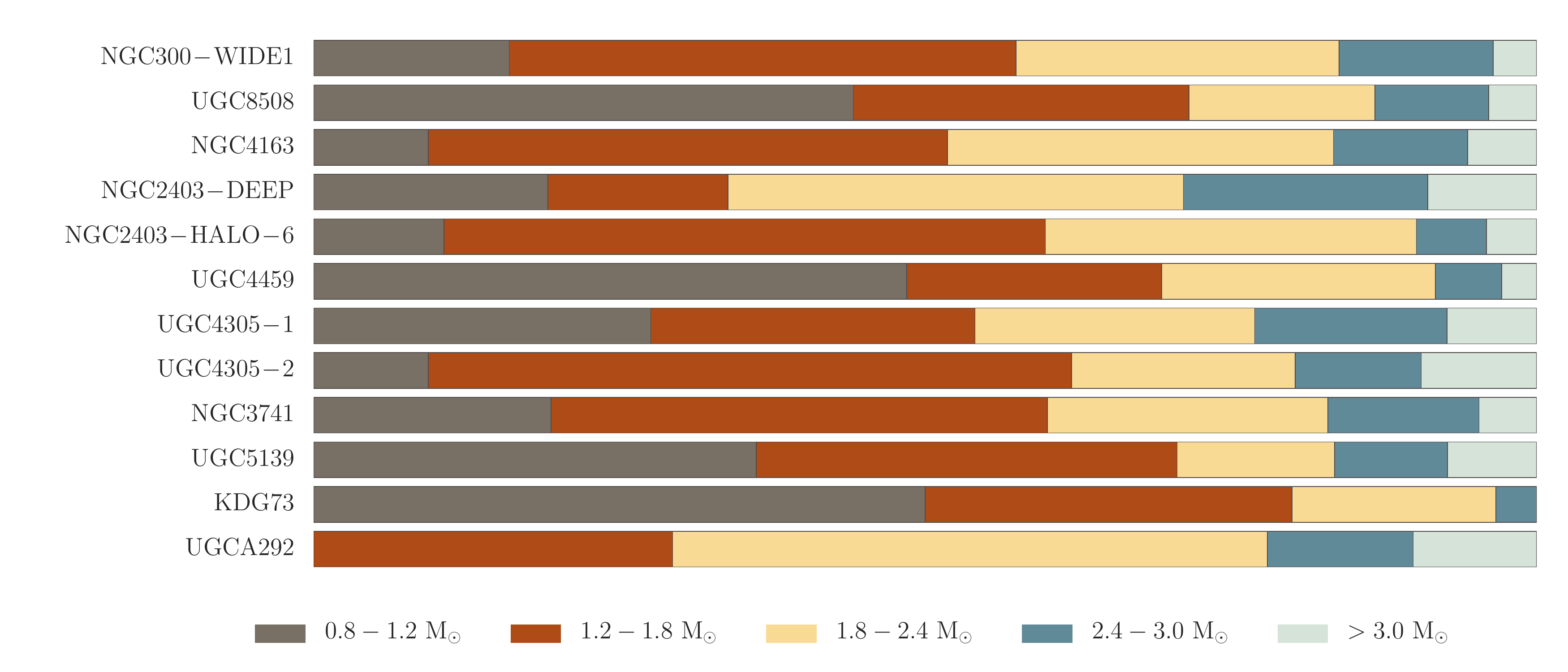}
\caption{Predicted TP-AGB age (top) and initial mass (bottom) distributions of each galaxy from the median values of the $\sim100$ {\tt TRILEGAL} simulations that sampled the random uncertainties of the derived SFH.}
\label{fig:tpagb_dist}
\end{figure*}

\begin{figure*}
\includegraphics[width=\textwidth]{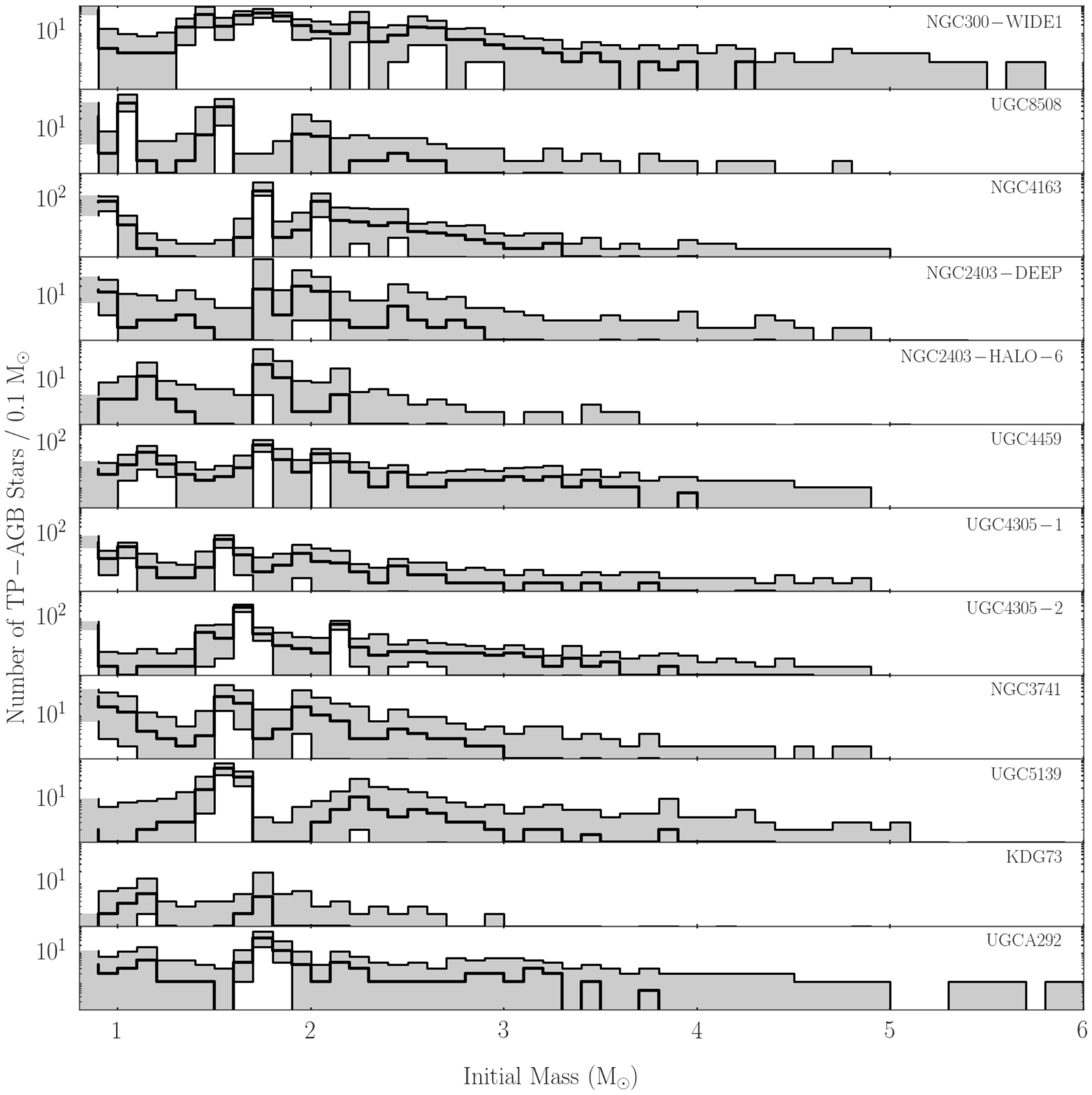}
\caption{Predicted TP-AGB mass distributions of each galaxy from the $\sim100$ {\tt TRILEGAL} simulations that sampled the random uncertainties of the derived SFH. Black histogram is the median LF plus gray dashed area for the 67\% confidence level interval of the combined simulations. Note masses $M\gtrsim3\Msun$ are TP-AGB stars that are expected to experience HBB.}
\label{fig:tpagb_masses}
\end{figure*}

\section{Conclusions}
\label{sec_conc}

The potential of using nearby galaxies with resolved SFHs to constrain TP-AGB evolution was demonstrated in our previous works (\citetalias{Girardi2010}, \citetalias{Rosenfield2014}), in which evolutionary parameters have been changed in order to improve the agreement between the model predictions and the data. The present work is intended to continue along this line with the addition of galaxies containing younger populations, hence allowing us to calibrate the parameters of TP-AGB stars with larger initial masses even into the regime of TP-AGB stars that experience HBB. We find that the models produced by \citetalias{Rosenfield2014} show good agreement in most cases, with a few exceptions. First, the \narratio\ ratio, a proxy for the mean TP-AGB lifetimes, showed a mean fractional difference data to model of 9\% across the entire galaxy sample. Second, we compared predicted and observed LFs, a more complex estimate of the TP-AGB lifetimes than the \narratio\ ratio. All but a few cases show consistent agreement to within random SFH uncertainties. Finally, we compared the predicted TP-AGB flux to the observed and found significant improvement compared to older TP-AGB models. The few problematic cases we find (KDG~73, UGCA~292) probably correspond to cases in which the SFH was not as well measured due to the depth of the photometry and/or the large distance to the galaxies.

We consider the present \citetalias{Rosenfield2014} models as satisfactory, and that the TP-AGB calibration process needs new and independent data to proceed, aiming to an even more stringent level of agreement between data and models.

The fact that the present models satisfactorily reproduce the numbers and LFs of the present HST optical and NIR data does not necessarily mean that they will equally well reproduce data in other passbands, even for the same galaxies. As discussed in \citetalias{Rosenfield2014}, the present calibration involved changes mainly in the mass-loss formula, as this was the largest parameter effecting the lifetime along the TP-AGB. There may be other ways of obtaining similar results via changes in other processes as well, including dredge-up efficiency, the mixing length parameter, along with a different mass-loss formalisms. These alternative ways, however, would probably imply different results for the mid-infrared photometry, which we cannot test with the present observational data for this sample.

A TP-AGB calibration following a multi-band and multi-parametric approach, applied simultaneously to many galaxies, is the clear path forward in understanding the physical properties of large numbers of TP-AGB stars. This process will be pursued in forthcoming works by our team, targeting more complete databases in the Magellanic Clouds and in M31, in addition to ANGST galaxies.

Present TP-AGB models are available at the STARKEY web server \url{http://starkey.astro.unipd.it}. The derived isochrones are provided via the \url{http://stev.oapd.inaf.it/cmd} and \url{http://starkey.astro.unipd.it/cmd} web interfaces as well as the regular grid of a subset of PARSEC V1.2S tracks used in {\tt MATCH}. The full grid of PARSEC V1.2S tracks are available at \url{http://people.sissa.it/~sbressan/parsec.html}

\acknowledgements
We acknowledge the support from the ERC Consolidator Grant funding scheme ({\em project STARKEY}, G.A. n.~615604),
and from the  from {\em Progetto di Ateneo 2012}, University of Padova, ID: CPDA125588/12.
All figures in this paper were produced using {\tt matplotlib} \citep{Hunter2007}. This research made use of {\tt Astropy}, a community-developed core Python package for Astronomy \cite{astropy}. This material is based upon work supported by the National Science Foundation under Award No. 1501205.

{\it Facilities:} \facility{HST (ACS)}, \facility{HST (WFPC2)}

\bibliography{tpagb2}

\end{document}